\RequirePackage{fix-cm}
\documentclass[smallextended]{svjour3}       
\smartqed  
\usepackage{float}
\usepackage{tabularx}
\usepackage{amsmath}
\usepackage{amsfonts}
\usepackage{algorithmic}
\usepackage[ruled,vlined, linesnumbered]{algorithm2e}
\usepackage{multirow}
\usepackage{graphicx}
\usepackage[export]{adjustbox}
\usepackage{caption}
\usepackage{nccmath}
\usepackage[strict]{changepage}
\usepackage{rotating}
\usepackage[labelformat=simple]{subcaption}

\usepackage{array,booktabs}
\usepackage{algorithm2e}
\usepackage{makecell}
\usepackage{tablefootnote}
\usepackage{threeparttable}
\usepackage{tcolorbox}
\usepackage{newtxtext,newtxmath}
\usepackage[numbers]{natbib}      
\bibpunct{[}{]}{,}{n}{}{,}         
\usepackage{enumitem}
\usepackage{fdsymbol}

\newcommand{\PreserveBackslash}[1]{\let\temp=\\#1\let\\=\temp}
\newcolumntype{C}[1]{>{\PreserveBackslash\centering}m{#1}}
\newcolumntype{R}[1]{>{\PreserveBackslash\raggedleft}m{#1}}
\newcolumntype{L}[1]{>{\PreserveBackslash\raggedright}m{#1}}
\begin{document}

\title{Search-based Selection of Metamorphic Relations for Optimized Robustness Testing of Large Language Models}


\author{Sangwon Hyun \and Shaukat Ali \and M. Ali Babar}


\institute{S. Hyun, M. A. Babar \at
              The University of Adelaide\\
              \email{dr.sangwon.hyun@gmail.com, ali.babar@adelaide.edu.au}           
           \and
           S. Ali \at
              Simula Research Laboratory \\
              \email{shaukat@simula.no}
           \and
}

\date{Received: date / Accepted: date}

\maketitle

\begin{abstract}
Assessing the trustworthiness of Large-Language Models (LLMs), such as robustness, has garnered significant attention.
Recently, metamorphic testing that defines Metamorphic Relations (MRs) has been widely applied to evaluate the robustness of LLM executions. 
However, the MR-based robustness testing still requires a scalable number of MRs, thereby necessitating the optimization of selecting MRs. Most extant LLM testing studies are limited to automatically generating test cases (i.e., MRs) to enhance failure detection. Additionally, most studies only considered a limited test space of single perturbation MRs in their evaluation of LLMs.
In contrast, our paper proposes a search-based approach for optimizing the MR groups to maximize failure detection and minimize the LLM execution cost. Moreover, our approach covers the combinatorial perturbations in MRs, facilitating the expansion of test space in the robustness assessment. 
We have developed a search process and implemented four search algorithms: Single-GA, NSGA-II, SPEA2, and MOEA/D with novel encoding to solve the MR selection problem in the LLM robustness testing. We conducted comparative experiments on the four search algorithms along with a random search, using two major LLMs with primary Text-to-Text tasks. Our statistical and empirical investigation revealed two key findings: (1) the MOEA/D algorithm performed the best in optimizing the MR space for LLM robustness testing, and (2) we identified silver bullet MRs for the LLM robustness testing, which demonstrated dominant capabilities in confusing LLMs across different Text-to-Text tasks.
In LLM robustness assessment, our research sheds light on the fundamental problem for optimized testing and provides insights into search-based solutions.
\keywords{Large-Language Model, Robustness Testing, Metamorphic Testing, Search-based Optimization}
\subclass{68N30 \and 68T20 \and 68M99}
\end{abstract}

\section{Introduction}
Testing Large-Language Models (LLMs) is critical to ensure their essential qualities, such as correctness, robustness, and AI ethics, in diverse application scenarios. While the correctness assessment has been prioritized by comparing LLM outcomes with benchmark datasets, recent studies~\cite{fan2023large, liu2020adversarial,jin2020bert,wang2023adversarial} have concentrated on evaluating the robustness of LLM executions, which significantly impacts their trustworthiness~\cite{liang2023holistic}. For example, fine-tuned LLMs used in digital healthcare, which access private patient notes~\cite{moor2023foundation, gao2021limitations}, must demonstrate high robustness to prevent potential data leakage issues.  Additionally, enterprise chatbots in the banking and education sectors~\cite{ribeiro2020beyond,okonkwo2021chatbots} should be validated to ensure feasible trustworthiness in practical use.

\begin{figure}[t]
    \centering
    \includegraphics[width=0.82\textwidth, trim = 0.7cm 0.4cm 0.4cm 0.3cm,clip]{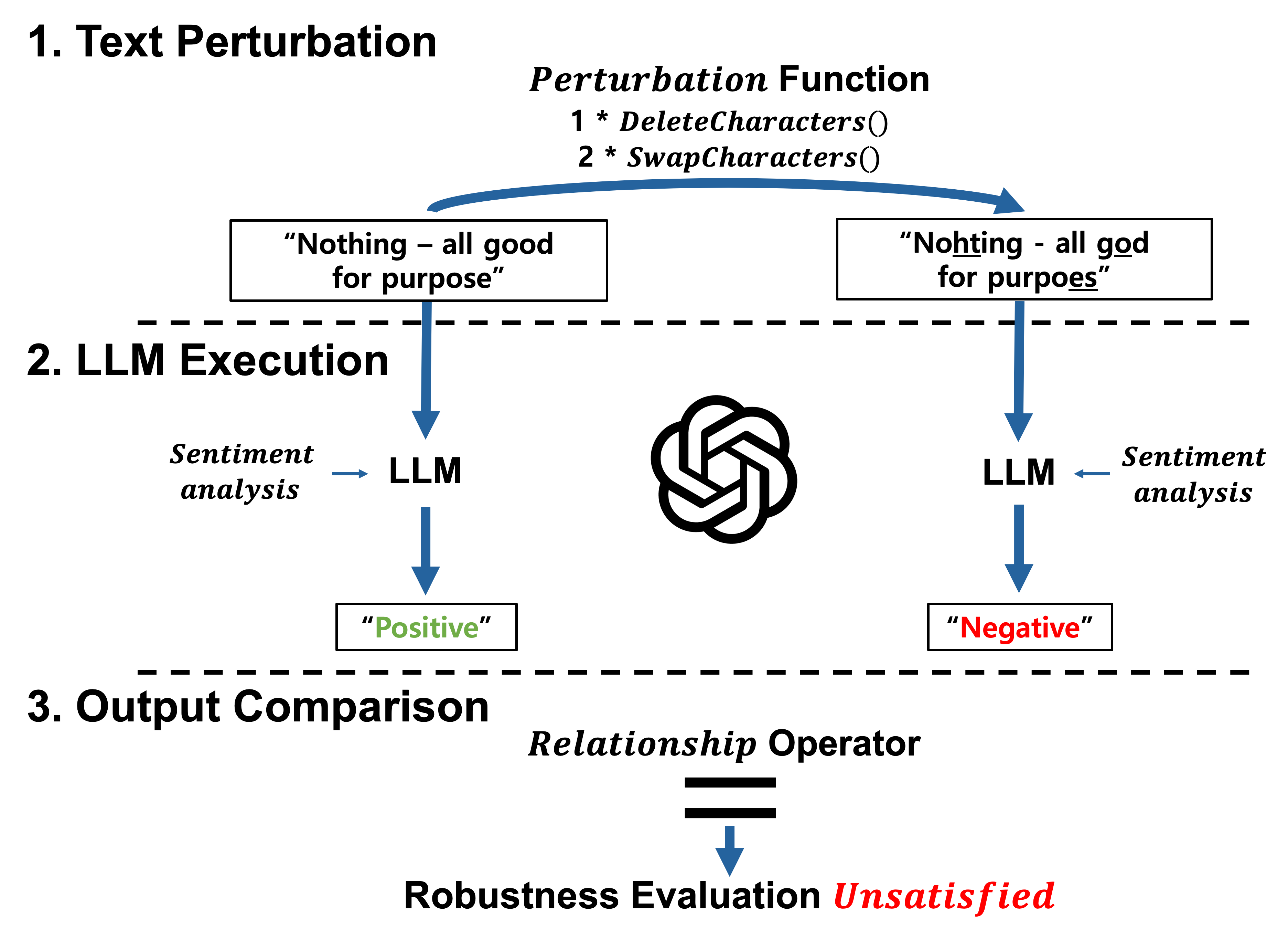}
    \caption{Metamorphic relation-based LLM robustness testing process with examples} \label{fig:intro}
    \vspace{-15px}
\end{figure}

The robustness testing aims to detect the altered outcomes of LLMs by adding intentional (e.g., adversarial attacks) or unintentional (e.g., typos) perturbations to the input text. Our previous study has defined this robustness testing process as a Metamorphic Relation (MR)-based scheme encompassing various perturbation methods and comparison metrics~\cite{hyun2024metal}. 
Fig.~\ref{fig:intro} describes an example evaluation process with MRs on LLMs. The process begins by transforming the input text, ``Nothing - All good for purpose", into follow-up texts, ``No\underline{ht}ing! - All g\underline{  }od for purpo\underline{es}", using perturbation functions, such as \textit{DeleteChracter() and \textit{SwapChracter()}}. Next, LLMs are executed with input texts and task instruction sentences. Finally, the two outputs generated by executing the original and follow-up text are compared by the relationship operator, like $=$. In the example, the LLM failed to satisfy the sentiment analysis on robustness evaluation because the original and follow-up outputs differ from ``Positive" and ``Negative." Each evaluation process is defined as an MR, serving the roles of test case and oracle for the LLM robustness assessment.

Even though the MR-based templates facilitate the pipelined execution of LLM robustness testing, they still incur substantial costs for running LLMs on the entire MR set to detect potential robustness failure cases~\cite{hyun2024metal,chiang2023can,wang2022semattack}. This challenge arises from the fact that (1) the space of MRs in robustness testing is virtually infinite and (2) the effectiveness of each MR can vary across different application scenarios~\cite{hyun2024metal, xu2024an}. The infinite MR space indicates that an input text, as shown in Fig.~\ref{fig:intro}, can be perturbed in unbounded ways, exemplified by ``Not\underline{i}ing - all good for \underline{pp}urpose," through the exponential combination of perturbation functions. Furthermore, existing studies have not observed any silver bullet MR that is universally effective for revealing robustness issues across different LLM application scenarios.

To address the challenges, we extended the fundamental MR selection problem, which identifies a minimum set of effective MRs for specific tasks~\cite{zheng2024identifying,chen2018metamorphic}, to the LLM robustness testing. The objective of the problem specification is as follows: 
\begin{center} \nonumber
    \textnormal{\textit{``The optimized selection of MR group while maintaining the capability of}} \\ \nonumber
    \textnormal{\textit{detecting failures and keeping the LLM execution costs low."}}
\end{center} 
Our study aims to search for an optimal subset of MRs, thereby enabling efficient and effective robustness testing to ensure the trustworthiness of LLMs in various application scenarios.

The investigation into LLM robustness testing and MR-based optimization studies~\cite{ayerdi2021generating, cho2022automatic,ebrahimi2018hotflip,hyun2024metal,jin2020bert,li2019textbugger,liu2020adversarial,polo2024tinybenchmarks,wang2023adversarial,zheng2020evaluating} revealed that most studies (1) have limited scope of defining new MRs (i.e., perturbation methods) for specific application scenarios and (2) have utilized particular single-type perturbations instead of combining multiple perturbations. 

First, all extant studies focused on automating the LLM quality evaluation by proposing new MRs (i.e., perturbation methods), except for the two studies~\cite{hyun2024metal,polo2024tinybenchmarks}. Our previous study~\cite{hyun2024metal} estimated the contribution of MRs for each evaluation. However, the results only provided individual MRs' contributions to specific assessments and did not investigate the effectiveness of MRs and their execution costs. Polo et al.~\cite{polo2024tinybenchmarks} proposed a benchmark sampling method for efficient LLM testing but still suffered from expandability issues due to the dependency on the oracle in the benchmark. A novel optimization approach is needed to test LLMs' robustness in diverse scenarios efficiently.

Second, most extant studies have designed and applied a single type of text perturbation in the LLM quality evaluation. However, several studies have addressed the effectiveness of multi-level perturbations (e.g., applying \textit{DeleteCharacter()} and \textit{ReplaceSynonym()} to the same text) in evaluating natural language tasks~\cite{ebrahimi2018hotflip,li2019textbugger,  zheng2020evaluating}. Furthermore, combined perturbations can expand the range of test space for LLM qualities beyond applying single-level perturbations, which increases the chances of finding unreported LLM quality issues. 

This study addressed the fundamental MR selection problem in the LLM robustness testing process by defining novel encodings for MR-based test space and proposing a search-based optimization process to scope cost-efficient and effective MR subsets. Our study also covers the test space of single and multi-level perturbations. We implemented the four conventionally utilized search algorithms based on the search process: Single-GA~\cite{mitchell1998introduction}, NSGA-II~\cite{deb2002fast}, SPEA2~\cite{zitzler2001spea2}, and MOEA/D~\cite{zhang2007moea}.

We conducted experiments to compare the four search algorithms with a random search on two primary Text-to-Text tasks using Gemini 1.5 pro~\cite{team2024gemini} and Llama 3.1 70B~\cite{grattafiori2024llama} models. Our findings reveal that (1) the MOEA/D algorithm achieved the best optimization fitness across all the experiment cases while the Single-GA achieved the most similar optimization performance, (2) the MOEA/D also required the most significant amount of execution overhead compared to the other algorithms, and (3) we identified the ``silver bullet" MRs on Text-to-Text tasks across different LLMs, particularly the graphical transformation and contextual synonym replacements, which demonstrated the dominant capabilities in confusing LLMs.

Our paper mainly contributes to the LLM robustness testing by 
\begin{itemize}
    \item Formulating an MR-based test space considering the combinatorial perturbations for the optimized MR selection,
    \item Pipelining a search process to select the most optimized MR groups with defining cost and effectiveness objectives in LLM robustness testing,
    \item Experimenting with primary search algorithms on robustness assessment of various experimental cases on Text-to-Text generations,
    \item Identifying dominant MRs across all the optimized outcomes, which can be employed to evaluate the robustness of LLMs in various tasks with high priority.
\end{itemize}

The paper is organized as follows: Section~\ref{sec.problem} explains the problem definition. Section~\ref{sec.appr} elaborates on the proposed approach. Sections~\ref{sec.expr} describe the experiment and empirical analysis results. Works related to this study are presented in Section~\ref{sec.related}. Finally, Section~\ref{sec.conclusion} concludes the study with directions for future work.

\section{Problem Definition: MR Selection in LLM Robustness Assessment} \label{sec.problem}

In metamorphic testing, an MR Group (MG) is applied to detect failures in target systems, substituting the role of a test suit~\cite{zheng2024identifying}. The selection process based on the given set of MRs is typically carried out by domain experts.
This study aimed to automate the selection process using search-based optimization methods to maximize the detection of LLM robustness issues while minimizing the cost required during LLM testing. In order to develop the search-based solutions, we have defined the test space of MRs covering the multi-level application of perturbations, the costs required for LLM robustness testing, and failure detection capabilities. 
We believe that we first formalized the token cost of LLM execution and combinatorial MR-based test space to optimize the MG for LLM robustness assessment.

We first defined the search space of MRs for evaluating LLM robustness. Based on the MR notation from our previous study~\cite{hyun2024metal}, we formalized a set of $MR$s, $Set\_MR$, given $Text \ni text$ where a $text$ indicates a text phrase, as follows:
\begin{flalign} 
    & Set\_MR = \{ MR \, | \,MR: Text \times Text \times P \to Bool \} \label{eq.MRs} \\
    & Perturb = \{  P\, |\, P:Text \to Text \} \label{eq.perturb}
\end{flalign}
An $MR$ can be interpreted as a function that returns a boolean value, indicating the satisfaction of the evaluation results. An $MR$ takes a task instruction text, an input text, and a perturbation function $P$ as described in Equation~(\ref{eq.MRs}). Equation~(\ref{eq.perturb}) defines the perturbation function $P$, which returns the perturbed text given the original text. For example, the task instruction text can be ``Please analyze the sentiment of the given text", the input text can be ``Nothing - all good for purpose.", the perturbation function can be \textit{DeleteChracter()}, and the outcome of the $MR$ will be True or False.

\begin{figure}[t]
    \centering
    \includegraphics[width=0.85\textwidth, trim = 0.6cm 0.8cm 0.5cm 0.4cm,clip]{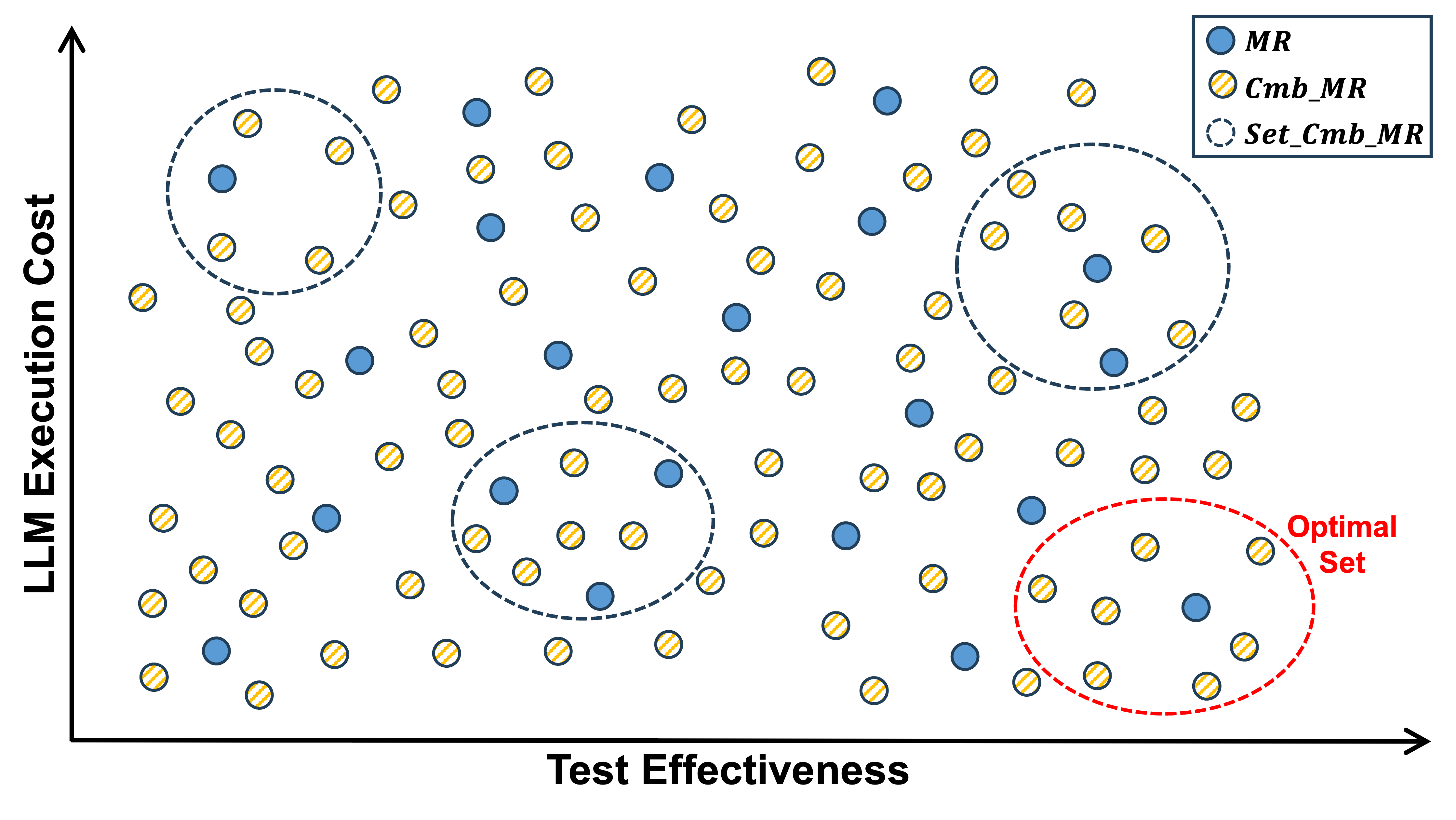}
    \caption{Example visualization of $MR$, $Cmb\_MR$, and $Set\_CmbMR$ spaces from the perspectives of LLM execution cost and testing effectiveness}\label{fig:problem}
    \vspace{-15px}
\end{figure}

Based on the definition from our previous study, we extend the space of $MR$s from single-level to multi-level (a.k.a combinatorial) perturbations to increase the $MR$ test space and improve the failure detection. To facilitate the combinatorial application of perturbations in a single $MR$, we defined $Cmb\_MR$ and $Set\_Cmb\_MR$ by extending $Set\_MR$ as follows:
\begin{flalign}
    & Set\_Cmb\_MR \triangleq \{ Cmb\_MR \, | \,Cmb\_MR: Text \times Text \times Cmb\_P \to Bool\}\label{eq.c.MRS} \\
    & Cmb\_P \triangleq (P_{i1} \circ ... \circ P_{ik}) \textnormal{, where } \{P_{i1}, ... P_{ik}\} \subset Perturb \textnormal{, and } \mathbb{N}_1^n \ni k \label{eq.c.p}
\end{flalign}
Equation~(\ref{eq.c.MRS}) represents a set of $MR$s encompassing both single and multiple applications of perturbation functions. $Cmb\_P$ in Equation~(\ref{eq.c.p}) indicates the $k$ iterative calls of perturbation functions, $P$, where the $\circ$ operator denotes function compositions. The indices $i1, i2, ..., ik$ in the subset of $Perturb$ refer to the positions of perturbation functions within a subset. 
For example, in the single $MR$, one type of perturbation function, $P$, such as \textit{DeleteCharacter()} or \textit{SwapCharacter()}, is applied three times in generating perturbed text for robustness testing. Conversely, the $Cmb\_MR$ refers to the use of different $P$s; for instance, applying \textit{DeleteCharacter()} twice and \textit{SwapCharacter()} once to generate a perturbed text from the original text.

{The example spaces of $MR$s, $Cmb\_MR$s, and $Set\_Cmb\_MR$s are visualized in Fig.~\ref{fig:problem}. The LLM execution cost reflects the necessary expenses, such as tokens or time, for testing the robustness of LLMs on specific tasks using corresponding $MR$s. Testing effectiveness refers to the number of revealed robustness failures while evaluating the LLMs on the tasks with the $MR$s. By considering the extended $Cmb\_MR$s, our study expands the search space compared to the single $MR$ spaces, thereby facilitating the search for the global optimal $Set\_Cmb\_MR$ for efficient and effective robustness assurance of LLMs on specialized tasks.}

Next, we have formulated two optimization objectives in this study: the efficiency and effectiveness of LLM robustness testing. For the efficiency factor, we defined the execution cost of LLMs by considering the number of word tokens in a text given to and generated by LLMs. Various metrics are used to measure the cost of executing LLMs, including time, number of queries, and number of tokens. We concluded that counting tokens is the most practical metric to evaluate LLM execution cost because the ChatGPT API uses the metric in its pricing policy~\cite{ChatGPT} and the other LLM services from major industry vendors have internal limits on tokens~\cite{Llama, chowdhery2022palm}. We define cost values, $\mathbb{R} \ni C\_token,\, C\_input\_token,\, C\_output$ $\_token$ for specifying the LLM execution cost as follows:
\begin{flalign}
    & C\_token = C\_input\_token + C\_output\_token \label{eq.c.total}\\
    & \{C\_input\_token,\; C\_output\_token\} \ni c = NumTokens(text) \label{eq.c}
\end{flalign}
In Equation~(\ref{eq.c.total}), $C\_token$ represents the total count of tokens that pass through LLMs. Equation~(\ref{eq.c}) defines that $C\_input\_token$ and $C\_output\_token$ can be calculated by calling $NumTokens: Text \to \mathbb{R}$ function, which returns the number of tokens present in the given input or output $text$s. 

Lastly, we estimated the failure detection capability (a.k.a test effectiveness) by using the Attack Success Rate ($ASR$) method, which is the most prevalent test effectiveness metric in LLM robustness assessment~\cite{zou2023universal, jin2020bert, li2019textbugger}. This metric is calculated by dividing the number of input texts that result in ``Unsatisfied" between original and perturbed outputs by the total number of input texts. For example, if the $MR$ in Fig.~\ref{fig:intro} makes 10 different results from 30 input texts, the ASR for the $MR$ is 0.33 (10/30). By computing the average $ASR$ value for each $MR$ in a selected MG, we can determine the effectiveness of the $MG$ in revealing robustness issues in LLMs. In addition, we consider the contextual similarity of the original input text and the perturbed text to accurately calculate the test effectiveness. The details of the cost and $ASR$ calculation methods are explained in Section~\ref{sec.appr.fitness}.

The defined equations are used to search for the optimal MG, a subset of the $Set\_Cmb\_MR$, for LLM robustness assessment which can achieve a maximum of $ASR$ and a minimum of $C\_token$ values. 

\section{Search-based Solutions}\label{sec.appr}

\begin{figure}[t]
    \centering
    \includegraphics[width=0.7\textwidth, trim = 0cm 0cm 0.5cm 0cm,clip]{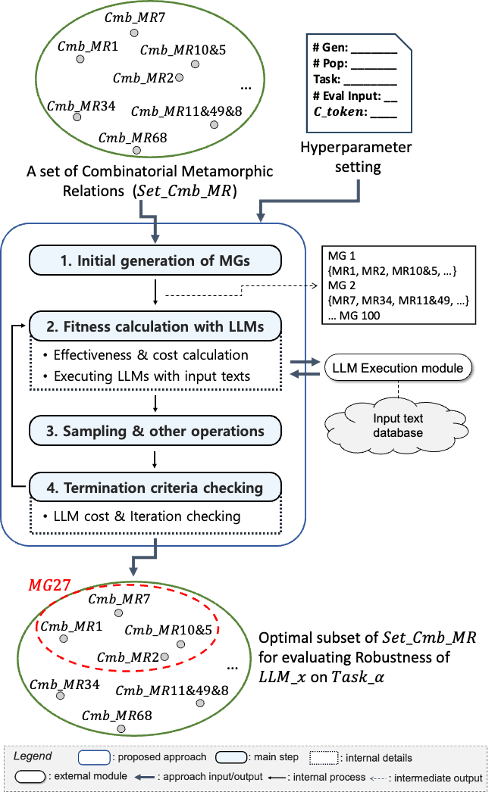}
    \caption{Overall execution process of the search-based optimization for MGs} \label{fig:overall}
    \vspace{-10px}
\end{figure}

This section explains the search-based solutions for the optimized selection from $Set\_Cmb\_MR$ by minimizing $C\_token$ and maximizing $ASR$ values. After introducing the overall process of search-based optimization for LLM quality evaluation, we will explain the four different search methods: Single-GA~\cite{mitchell1998introduction}, NSGA-II~\cite{deb2002fast}, SPEA2~\cite{zitzler2001spea2}, and MOEA/D~\cite{zhang2007moea}, which are implemented in our study.

\subsection{Overall Search Process} \label{sec.appr.overall}
Fig.~\ref{fig:overall} describes the overall execution process of the search algorithms based on the problem definition in Section~\ref{sec.problem}. This approach receives two main inputs: a pre-defined set of $MR$s to evaluate LLM robustness and a set of hyperparameter settings, such as maximum generations, number of $MG$s, and maximum tokens for LLM execution cost, $C\_token$. The hyperparameter settings in the experiment are described in Section~\ref{sec.expr.design}.

Using the two sets of inputs, the search process consists of four steps: generating an initial set of MR Groups ($MG$s), calculating fitness and cost with LLM execution, executing sampling operations, including selection and mutation in GAs, and checking termination criteria. The second to last steps are iterated until the optimal $MG$, a subset of $Set\_Cmb\_MR$, is searched, or the termination criteria are met. This study also introduces the novel encodings for defining $MG$s and sampling operators in the four GA implementations. Finally, the process returns the last $MG$, considered the optimized subset of $Set\_Cmb\_MR$ for evaluating the LLM robustness on target tasks. 

\subsection{Initial Generation of Subsets} \label{sec.appr.MRs}

\begin{figure}[t]
    \centering
    \includegraphics[width=0.9\textwidth, trim = 0cm 0.2cm 0cm 0.3cm,clip]{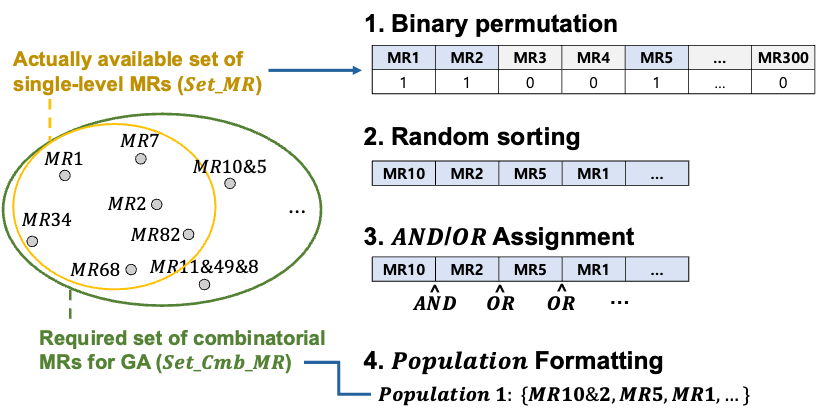}
    \caption{Example execution of the initial $MG$ generation module for $Cmb\_MR$s} 
    \label{fig:appr.pop}
    \vspace{-10px}
\end{figure}

The proposed approach assumed the existence of pre-defined $MR$s or similar adversarial attack generation models~\cite{wang2022semattack,chiang2023can} that can be represented to the $Cmb\_MR$ function format defined in Equation (\ref{eq.c.MRS}). Although some sets of single-level $MR$s are available for LLM quality evaluation~\cite{hyun2024metal}, we believe no executable sets of $Cmb\_MR$s have been released in existing studies or repositories. Therefore, we developed the $MG$ generation module that uses the available sets of single-level $MR$s to randomly generate initial groups of $Cmb\_MR$s for the search process. 

We defined a $MG$ as a subset of $Set\_Cmb\_MR$ in the proposed search process:
\begin{flalign}
    \nonumber
    & Set\_Cmb\_MR \supset MG = \{ Cmb\_MR \, | \, Cmb\_MR\, \textnormal{ from }\\
    & \qquad \qquad \qquad \qquad \qquad \,\, CombGen: Set\_MR \to Cmb\_MR\} \label{eq.pop}
\end{flalign}
The initial $MG$ generation module aims to create candidate $MG$s by utilizing a set of single-level $MR$s, $Set\_MR$. To bridge the gap between the available $Set\_MR$ and the required $Set\_Cmb\_MR$ in an efficient manner, we have proposed a binary permutation-based $MG$ generation method, $CombGen$ in Equation~(\ref{eq.pop}), which can generate a $Cmb\_MR$ from $Set\_MR$. As described in Fig.~\ref{fig:appr.pop}, the first step is randomly assigning binary values to each available single-level $MR$s. Then, the selected $MR$s are randomly sorted. Next, the $AND$ and $OR$ operators are randomly assigned, where $AND$ represents the combination of $MR$s and $OR$ indicates the independent $MR$. Finally, the module returns a $MG$ covering the single and combinatorial $MR$s. These steps are repeated until a specified number of $MG$s are generated.

For example, based on Fig.~\ref{fig:appr.pop}, assuming that $MR10$ applies the \textit{DeleteCharacter()} perturbation function, while $MR2$ utilizes the \textit{SwapCharacter()} function, the $MR10\&2$ generated by the $CombGen$ module signifies the combination of these two perturbations. Each perturbation method is executed as a separate function in our framework, meaning that $MR10\&2$ runs both functions to produce perturbed inputs from the original text. On the other hand, the $OR$ operator in the $CombGen$ module marks a dividing point for each generated $Cmb\_MR$. For example, if $MR5$ employs the \textit{ReplaceCharacter()} perturbation and has $OR$ operators on both sides, it is treated as an independent $Cmb\_MR$ that solely carries out this single perturbation function.

The initial $MG$ generation module facilitates the efficient extension of single-level $Set\_MR$ to the geometrically expanding search space of $Set\_Cmb\_MR$. 

\subsection{Fitness Calculation with LLMs} \label{sec.appr.fitness} 

The iterative process begins after the initial $MG$s are generated and involves the fitness calculation, sampling, and termination checking steps. We have defined a fitness function that considers test effectiveness and LLM execution costs by using the Attack Success Rate ($ASR$) and the input/output token costs of LLMs ($C\_token$).


\textbf{Effectiveness Calculation Considering Perturbation Qualities.} As described in Section~\ref{sec.problem}, this study uses the average $ASR$ of the constituent $Cmb\_MR$s in an $MG$ to quantify the test effectiveness the generated groups. However, a significant risk exists in directly applying the $ASR$ metric: contextual consistency between the original and perturbed texts. For example, the input text in Fig.~\ref{fig:intro} can be perturbed to ``Nohin! - All od fr purpe" by several number of executions on \textit{DeleteCharacter()} function. The example perturbed text has a significant contextual difference from the original input text: ``Nothing - All good for purpose." These cases should be regarded as \textit{false-error} test cases that adversely affect the confidence of LLM robustness assessment results. 

Therefore, we introduced the $Context\_ASR$ metric to encompass the original $ASR$ and the contextual difference values for improved confidence in the assessment results. We employed the $PerturbationQuality$ metric, which measures the similarity of vector encodings of input and perturbed texts, as suggested in our prior research~\cite{hyun2024metal}. The $Context\_ASR$ is calculated by multiplying $ASR$ and $PerturbationQuality$ values. For example, the $Cmb\_MR$ example in Fig.~\ref{fig:intro} may have 0.33 $ASR$ and 0.87 $PerturbationQuality$ values. In another \textit{false-error} example provided in this section, $ASR$ may be 0.68 $ASR$ while $PerturbationQuality$ is 0.11. We can then prioritize the non \textit{false-error} $Cmb\_MR$s by comparing the 0.29 and 0.07 $Context\_ASR$ values, respectively. We utilized Google's USE model for text encoding~\cite{cer2018universal} and cosine similarity for comparing encoded vectors to calculate the $PerturbationQuality$ values. The details of text effectiveness analysis for context-preserving and context-altering perturbations are described in Section~\ref{sec.expr.tv}.

\textbf{Cost Calculation.} As defined in Equations~(\ref{eq.c.total}) and (\ref{eq.c}), we have opted to use the number of communicated tokens with LLMs, $C\_token$, as our cost estimation metric. We count the tokens of LLMs' input and outputs during the $MR$ effectiveness calculation using the tokenizer provided by ChatGPT~\cite{GPTTokenizer}.

Algorithm~\ref{alg:effec_cost} explains the pseudo-code of calculating the test effectiveness and token cost needed to execute LLMs for the evaluation. Given an $MG$, a set of random input text, $eval\_input\_text$, and an instructive text for ordering specific task, $task\_instruction$, the algorithm returns the $Context\_ASR$ and $C\_token$ values. 

After initializing the lists in Line 1, the nested loop for iterating a pair of $Cmb\_MR$ and $input$ text is elucidated from Lines 2 to 11. To minimize the overhead of calculating fitness values, we developed a caching-based LLM execution module. Lines 4 to 5 denote the cache reading codes if the pair of $Cmb\_MR$ and $input$ hits the cache log. Otherwise, the $output$ and $eval\_result$ are generated by calling $ExecuteLLM$, $ASR$, and $PerturbationQuality$ functions, followed by the cache writing described in Lines 6 to 9. In the implementation, we utilized the LLM execution module and $ASR$, $PerturbationQuality$ calculation functions provided by METAL~\cite{hyun2024metal}. Finally, we calculate the $Context\_ASR$ and $C\_token$ using the saved $EvalResults$ and $TokensCount$ for each pair of $Cmb\_MR$ and $input$ in Lines 12 and 13. The implementation detail is opened in the repository\footnote{https://zenodo.org/records/15205020}. 

In our implementation of the four search algorithms based on JMetalPy framework~\cite{benitez2019jmetalpy}, we utilized the defined two methods to calculate the fitness of searching the optimal $MG$ for the LLM robustness testing. For Single-GA, we defined the following fitness function to encompass the two objectives of effectiveness and efficiency:
\begin{flalign} \nonumber
    Fitness\_Single = w_1 * (1-Context\_ASR) + w_2 * Normalization(\,C\_token\,).
\end{flalign}
We defined $1-Context\_ASR$ to be compatible with the minimizing optimization of the JMetalPy framework. Additionally, we normalized the $C\_token$ values according to the input and expected output token ranges for each task in our experiment. The $Context\_ASR$ values were not normalized because they already have a specific range from 0 to 1.
The weights of each objective are equally assigned as 0.5 in our study, but this parameter setting can be adjusted depending on the priority of a specific objective.

In multi-objective algorithms, including NSGA-II, SPEA2, and MOEA/D, we assigned the two fitness functions as follows:
\begin{itemize}[label={\(\smallblackcircle\)}, leftmargin=100px]
    \item $Fitness_1 = 1-Context\_ASR$
    \item $Fitness_2 = Normalization(C\_token)$.
\end{itemize}
The fitness functions are implemented in the EquationProblem.py file for each multi-objective search class in the JMetalPy framework.

\SetKwInOut{Parameter}{Parameter}
\SetAlgoNoEnd
\setlength{\textfloatsep}{0pt}
\begin{algorithm}[t!] 
\SetAlgoLined
\SetKwInOut{Input}{Input}
\SetKwInOut{Output}{Output}
 \caption{Effectiveness and Cost Calculation for Fitness} \label{alg:effec_cost}
\Input{$MG \subset Set\_Cmb\_MR$, $eval\_inputs \subset Text$ $task\_instruction \in Text$}
\Output{$Context\_ASR,\; C\_token \in \mathbb{R}$}
$EvalResults, TokensCount, Cache \leftarrow []$\;
         \For{$Cmb\_MR$ in $MG$} {
            \For{$input$ in $eval\_input\_text$} {
              \eIf{($Cmb\_MR$,$input$) in $Cache$} {
                 $output$, $eval\_result$ = $ReadCache(Cmb\_MR,input)$\;
              } 
              {
                $output$ = $ExecuteLLM(input, task\_instruction)$\;
                $eval\_result$ = $ASR(input, output, Cmb\_MR)$ $*$ $PerturbationQuality(input, Cmb\_MR)$\;
                $WriteCache((Cmb\_MR,input), output, eval\_result)$\;
              }
            }
            $EvalResults.append(eval\_result)$\;
            $TokensCount.append(Token(input) + Token(output))$\;
        }
$Context\_ASR = Avg(EvalResults)$\;
$C\_token = Sum(TokensCount)$\;
return~\parbox[t]{1\linewidth}{$Context\_ASR$, $C\_token$}
\end{algorithm}

\subsection{Sampling with GA Operations} \label{sec.appr.operations}
This section explains sampling methods that optimize search results based on fitness calculations. We introduce the sampling operations that can be applied to the general GA algorithms. Particularly, we have developed new crossover and mutation operators for the $MG$s consisting of $Cmb\_MR$s.

\textbf{Selection.} In the previous section, we explained two objectives: test effectiveness and token cost, and discussed two ways of utilizing them as fitness values in Single-GA and other multi-objective algorithms. For the selection methods, we applied the widely used binary tournament selection method for all the GA implementations~\cite{benitez2019jmetalpy}.

\begin{figure}[t]
     \centering
     \begin{subfigure}[b]{0.8\textwidth}
         \includegraphics[width=\textwidth, trim = 0cm 0cm 0cm 0cm,clip]{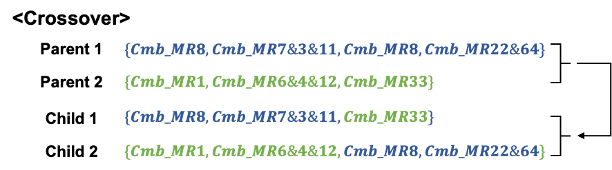}
         \caption{Example execution of crossover operation}
         \label{fig:exp-1}
         \vspace{5px}
     \end{subfigure}
     \begin{subfigure}[b]{0.8\textwidth}
         \centering
         \includegraphics[width=\textwidth, trim = 0cm 0cm 0cm 0cm,clip]{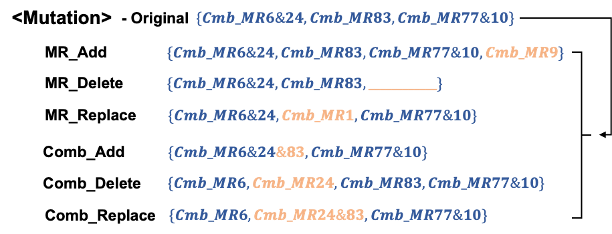}
         \caption{Example execution of mutation operation}
         \label{fig:exp-2}
     \end{subfigure}
     \caption{Examples of crossover and mutation operations}
     \label{fig:exp}
     \vspace{5px}
\end{figure}

\textbf{Crossover.} The crossover operation generates new child $MG$s that preserve the components in the best $MG$s in the previous generation. Fig.~\ref{fig:exp-1} illustrates the crossover methods proposed in this study. After selecting the most fit $MG$s as parents, the crossover operation creates child $MG$s by mixing the constituent $MR$s from these parents. Specifically, crossover points are randomly determined for each of the parent $MG$s. The first child $MG$ is formed by combining the initial components of the first parent with the latter components of the second parent, based on the division point. Conversely, the second child $MG$ is generated by taking the remaining components from both parents. As this study does not consider the execution order of $MR$s, we focused on conducting the one-point crossover operations instead of predicting the optimal crossover points.

\textbf{Mutation.} The mutation operation is crucial in the GA process to increase diversity and prevent the results from converging to a local optimal. Conventionally, three types of mutation operators- Add, Delete, and Replace - are used in the GA process~\cite{cho2022automatic,mitchell1998introduction}. However, because the $MG$s defined in our study consist of $Cmb\_MR$s, we defined $MR$ and combinatorial-level mutation operators by extending the three existing operators.

The top three examples in Fig.~\ref{fig:exp-2} illustrate the $MR$-level Add, Delete, and Replace operators. These operators randomly add new $Cmb\_MR$ into the original $MG$, delete a random $Cmb\_MR$ from the $MG$, and replace a random $Cmb\_MR$ with a new $Cmb\_MR$. The first example of MR\_Add, in Fig.~\ref{fig:exp-2}, shows the addition of new $Cmb\_MR9$ to the original $MG$. The second example demonstrates the deletion of the last component $Cmb\_MR77\&10$, while the third example describes the replacement of $Cmb\_MR83$ with $Cmb\_MR1$. 

The combination-level mutation operators are highlighted in the bottom three examples in Fig.~\ref{fig:exp-2}. To handle the unique features of $Cmb\_MR$s proposed in this study, we defined three more mutation operators for adding, deleting, and replacing $Cmb\_MR$s in their combinatorial notation. For example, the Comb\_Add operator adds the new $Cmb\_MR83$ to the first $Cmb\_MR6\&24$. The Comb\_Delete example illustrates the deletion of the combinatorial relationship between the $Cmb\_MR6$ and $Cmb\_MR24$, which decouples a $Cmb\_MR$ into two separate $Cmb\_MR$s. Finally, the Comb\_Replace operator is the consecutive application of Comb\_Delete and Comb\_Add operators. As seen in the last example in Fig.~\ref{fig:exp-2}, $Cmb\_MR24$ is divided from the $Cmb\_MR6\&24$, and $Cmb\_MR83$ is appended to $Cmb\_MR24$. The proposed mutation operators are randomly applied by following the uniform probability distribution~\cite{cho2022automatic}. The probability of mutation rate is explained in Section~\ref{sec.expr.design}.

\subsection{Termination Criteria}
In our search process, the final step is to check the termination criteria of the iterative execution. The main parameter for the termination criteria is the maximum number of iterations, followed by the difference threshold of the fitness values for two consecutive generations. The outcome of the search algorithms is the Elite Archive (i.e., dynamic Hall-of-Fame in JMetalPy) for Single-GA and Pareto Front solutions (i.e., non-dominated solution archive in JMetalPy) for multi-objective algorithms, which consist of sets of $Cmb\_MR$s identified during the executions.

\section{Experiment}
\label{sec.expr}
Our evaluation involved comparing the effectiveness and efficiency of the outcomes from the four GA algorithms and the random search algorithm by addressing three research questions outlined in Section~\ref{sec.expr.design}.

\newcommand\Tstrut{\rule{0pt}{2.5ex}}       
\newcommand\Bstrut{\rule[-1.3ex]{0pt}{0pt}} 
\newcommand\TBstrut{\Tstrut\Bstrut}         

\begin{table}[t]
\centering
\caption{Overall parameter setting and configuration for the experiment} \label{tab:setup}
\resizebox{0.70\textwidth}{!}{
\begin{tabular}{ll}
\hline \hline
\multicolumn{2}{c}{\TBstrut \normalsize{\textbf{GA Parameter Setting}}}  \\ \hline
 Population size                     & \Tstrut 100 \\ 
 Crossover rate              & \Tstrut 0.6  \\ 
 Mutation rate                   & \Tstrut 0.3 \\  
 Mutation operator execution rate  & \Tstrut 0.17 for six operators\\ 
 Maximum iterations               & \Tstrut 2,000  \\ 
 \# of single-level $MR$s         & \Tstrut 240 \\
 \# of $MR$s in a $MG$ & \Tstrut 3 to 30 \\
 \# of perturbation functions in $MR$s & \TBstrut 1 to 4 \\ \hline 
\multicolumn{2}{c}{\TBstrut \normalsize{\textbf{LLM Execution Setting}}}  \\ \hline
 Target LLMs & \Tstrut \begin{tabular}[c]{@{}l@{}} Gemini 1.5 Pro,\\ Llama 3.1 70B\end{tabular} \\
 Target quality attribute             & \TBstrut Robustness  \\ 
 Target LLM tasks                & \Tstrut \begin{tabular}[c]{@{}l@{}} Sentiment Analysis (SA),\\ Text summarization (TS)\end{tabular}   \\ 
 \Tstrut \begin{tabular}[c]{@{}l@{}} \# of evaluation input \\ texts per iteration \end{tabular}
            &  \begin{tabular}[c]{@{}l@{}} 50 random texts out of \\ 600 real-world texts \end{tabular}  \Bstrut \\ \hline
\multicolumn{2}{c}{ \TBstrut \textbf{Execution Environment}}                                                                                                                           \\ \hline
Python version                  & \Tstrut 3.11.4                                                                                                                             \\ 
Virtual environment             &  \Tstrut Conda 23.7.2                                                                                                                    \\ 
Memory                          & \TBstrut 16 GB                                                                                                                              \\ \hline \hline
\end{tabular}
}
\vspace{5px}
\end{table}

\subsection{Experiment Design} \label{sec.expr.design}

\textbf{Experiment Setting.} In executing search-based solutions, the population size (i.e., the number of $MG$s) for each iteration is set to 100, and the crossover and mutation rates are set to 0.6 and 0.3, respectively. During the mutation execution, each of the six operators defined in Section~\ref{sec.appr.operations} is applied with a uniform probability of 0.17. The maximum number of iterations is set to 2,000. Additionally, given to the GA process for the input set of single-level $MR$s, $Set\_MR$, we used 240 $MR$s generated by 42 perturbation functions in the METAL framework~\cite{hyun2024metal}. Finally, to alleviate the generation of outlier $MG$s, such as a $MG$ with a hundred number of $Cmb\_MR$s, we have set upper and lower limits for the number of $Cmb\_MR$s in a $MG$ and the number of single-level $MR$s in a $Cmb\_MR$. We maintained 3 to 30 independent $Cmb\_MR$s in a $MG$ and 1 to 4 single-level $MR$s in a $Cmb\_MR$ to limit the depth of combinations. The details of the empirical parameter settings are explained in Section~\ref{sec.expr.tv}. 

Next, this experiment evaluated the robustness of two primary tasks utilizing LLMs for text data: Sentiment Analysis (SA) and Text Summarization (TS). SA and TS are the representative tasks of the classification and generative task groups for LLMs~\cite{qiu2022adversarial, liang2023holistic}. 
The SA is a widely used task in natural language processing and falls under the classification task group. Similarly, TS is one of the most commonly applied scenarios for LLMs within the generative task group. We selected these two tasks because they represent conventional application scenarios for evaluating various quality attributes of LLM executions using different token distributions of text data~\cite{liang2023holistic,jin2020bert,zhu2023promptbench}. Furthermore, they are often implicitly utilized in other types of LLM tasks~\cite{liang2023holistic}. For example, when we pose specific questions to LLMs, they summarize the relevant text retrieved from sources and prioritize key points based on the user's context and the sentiment (i.e., temperature) of the explanation.

We used Google's Gemini 1.5 Pro~\cite{team2024gemini} and Meta's Llama 3.1 70B~\cite{grattafiori2024llama} models as target LLMs for assessing the robustness in executing selected tasks. To calculate the test effectiveness and token costs for each $MG$, we selected 50 input texts randomly. These input texts were chosen from 600 real-world text datasets from Amazon reviews and ABC news provided by METAL framework~\cite{hyun2024metal}.

Finally, the execution environment is explained for the reproduction guideline. We executed the experiment group in Python version 3.11.4 with a Conda 23.7.2 virtual environment, assigning 16GB of memory for the execution.

\textbf{Experiment Group.} This study implemented four genetic algorithm techniques to search for the optimal model generation for the robustness assessment of large language models, utilizing the JMetalPy framework~\cite{benitez2019jmetalpy}. Additionally, we developed a random search algorithm as a baseline experimental group in our study.

\makeatletter 
\g@addto@macro{\@algocf@init}{\SetKwInOut{Parameter}{Parameter}} 
\makeatother

\SetAlgoNoEnd
\begin{algorithm}[t!] 
\SetAlgoLined
\SetKwInOut{Input}{Input}
\SetKwInOut{Output}{Output}
 \caption{Random Search for Optimal $MG$} \label{alg:random_search}
\Input{$Set\_MR$, $total\_input\_text \subset Text$, $task\_instruction \in Text$}
\Parameter{$max\_iterations, num\_eval\_inputs, num\_MG \in \mathbb{N}$}
\Output{$Opt\_MG \subset Set\_Cmb\_MR$}
$best\_MG, next\_MGs, eval\_input\_text \leftarrow []$\;
$num\_iteration, sum\_costs \leftarrow 0$\;
         \While{$True$} {
              $eval\_inputs = randomInput(num\_eval\_inputs, total\_input\_set)$\;
              \eIf{$best\_MG$ is $Empty$} {
                $best\_MG.append(InitMG(Set\_MR))$\;
                $best\_effectiveness, best\_costs = Fitness(best\_MG)$\;
                $sum\_costs\, += best\_costs$\;
              } 
              {
                \For {$i$ in range($num\_MG$)} {
                $next\_MGs.append(InitMG(Set\_MR))$\;
                }
                \For {$i$ in range($num\_MG$)} {
                $next\_effectiveness, next\_costs = Fitness(next\_MGs[i])$\;
                $sum\_costs\, += next\_costs$\;
                \If {$best\_effectiveness < next\_effectiveness$} {
                  $best\_MG = next\_MGs[i]$\;
                  $best\_effectiveness = next\_effectiveness$\;
                  $best\_costs = next\_costs$\;
                }
                }
              }
              \If{$num\_iteration > max\_iterations$} 
              {Break\; }
        }
return~\parbox[t]{1\linewidth}{$best\_gen$, $best\_effectiveness$, $num\_tokens$}
\end{algorithm}

We have extended the common heuristics of the random search~\cite{harman2008search,zabinsky2009random} to solve the optimal $MG$ problem for assessing LLM robustness. Algorithm~\ref{alg:random_search} presents the pseudo-code for the random search in this experiment. This algorithm's inputs, parameters, and outputs are the same as the GA process explained in Section~\ref{sec.appr.overall}.

The proposed random search is a process to compare the $best\_MG$ with newly generated $MG$s, called $next\_MGs$. In each iteration, $next\_MGs$ has 100 randomly generated $MG$s. The algorithm updates the best generation whenever a new $MG$ achieves higher fitness values. This process is iterated until the maximum number of iterations is reached. The iterative process is described in lines 3 to 21 of Algorithm~\ref{alg:random_search}. In line 4, the evaluation set of input texts is randomly selected from the whole set of texts. Lines 5 to 8 show the first generation case, where the $best\_MG$ is empty. Each $MG$ is generated by executing \textit{InitMG()} with the initially given single $MR$ set, $Set\_MR$, as described in Section~\ref{sec.appr.MRs}. Then, the effectiveness and cost values for each generation are calculated using the $Fitness()$ function, explained in Algorithm~\ref{alg:effec_cost} in Section~\ref{sec.appr.fitness}. In the subsequent iterations shown in Lines 9 to 18, the $best\_MG$ is updated by comparing its fitness value with the ones of $MG$ in the $next\_MGs$. Finally, the algorithm checks the termination criteria using $num\_iterations$ in line 19 and returns the best generation, effectiveness, and cost values. 

The developed GAs and the random search were executed using the same LLMs, input texts, and parameter settings. We have evaluated the robustness of the SA and TS tasks for the four search algorithms and one baseline method. 
To account for the randomness in the algorithms, each algorithm execution was repeated 30 times. 

We have defined the following Research Questions (RQs) to accurately evaluate the efficacy and efficiency of the generated results from the experiment group:
\begin{itemize}
    \item RQ1. Do the proposed GAs present significantly better effectiveness and efficiency than the random search?
    \item RQ2. Which search-based algorithm yields the highest optimization performance?
    \begin{itemize}
        \item RQ2-1. Which search algorithm achieves the best effectiveness and efficiency?
        \item RQ2-2. Which search algorithm experiences significant execution overhead?
    \end{itemize}
    \item RQ3. Which $Cmb\_MR$s demonstrate significant testing capabilities for assessing the robustness of LLM text outcomes?
\end{itemize}

\textbf{RQ1.} The primary objective of RQ1 is to assess the effectiveness and efficiency of the proposed GA techniques compared to random search in the robustness evaluation of the target LLM. We conducted a comparative analysis of the outcomes of the two GA implementations and the random search algorithm.

\textbf{RQ2.} In this RQ, we thoroughly examined the search outcomes from the proposed four GA methods. We analyzed the HyperVolume indicator to determine which multi-objective algorithms yield the best search results. Furthermore, we compared the optimal multi-objective algorithm with the Single-GA approach. Additionally, we assessed the execution overhead of each algorithm to evaluate the practical feasibility of their application.

\textbf{RQ3.} In the last RQ, we empirically analyzed the outcomes of the proposed search algorithms and examined the dominant contributions and patterns of specific $MR$s and perturbation functions across different tasks and LLMs.

\subsection{Experiment Results} \label{sec.expr.result}

Our experiment generated 18,000,000 $MG$s, calculated as 100 $MG$s for each of the 1,200 maximum iterations multiplied by 30 repetitions across 5 algorithms. Additionally, we produced 55,732 independent $ Cmb_MR$s out of a total of 94,062 $ Cmb_MR$s, using 240 single-level $MR$s. This process also resulted in creating 1.2 GB of cache logs from LLM executions, utilizing 600 input texts.

\begin{figure}[t]
    \includegraphics[width=0.98\textwidth, trim = 0cm 0cm 0cm 0cm,clip]{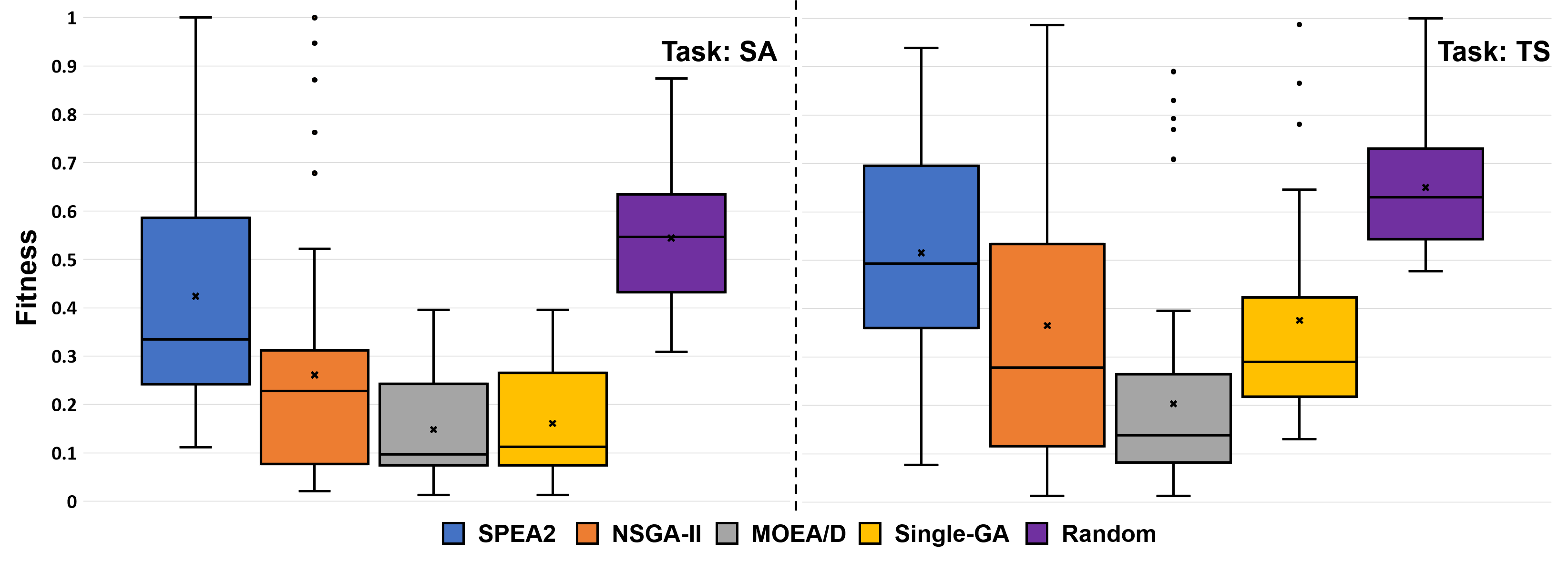}
    \caption{Fitness values from the experiment groups on evaluating the Robustness of SA and TS tasks for Gemini 1.5 Pro} \label{fig:RQ1-1}
    \vspace{5px}
\end{figure}

\begin{figure}[t]
    \includegraphics[width=0.98\textwidth, trim = 0cm 0cm 0cm 0cm,clip]{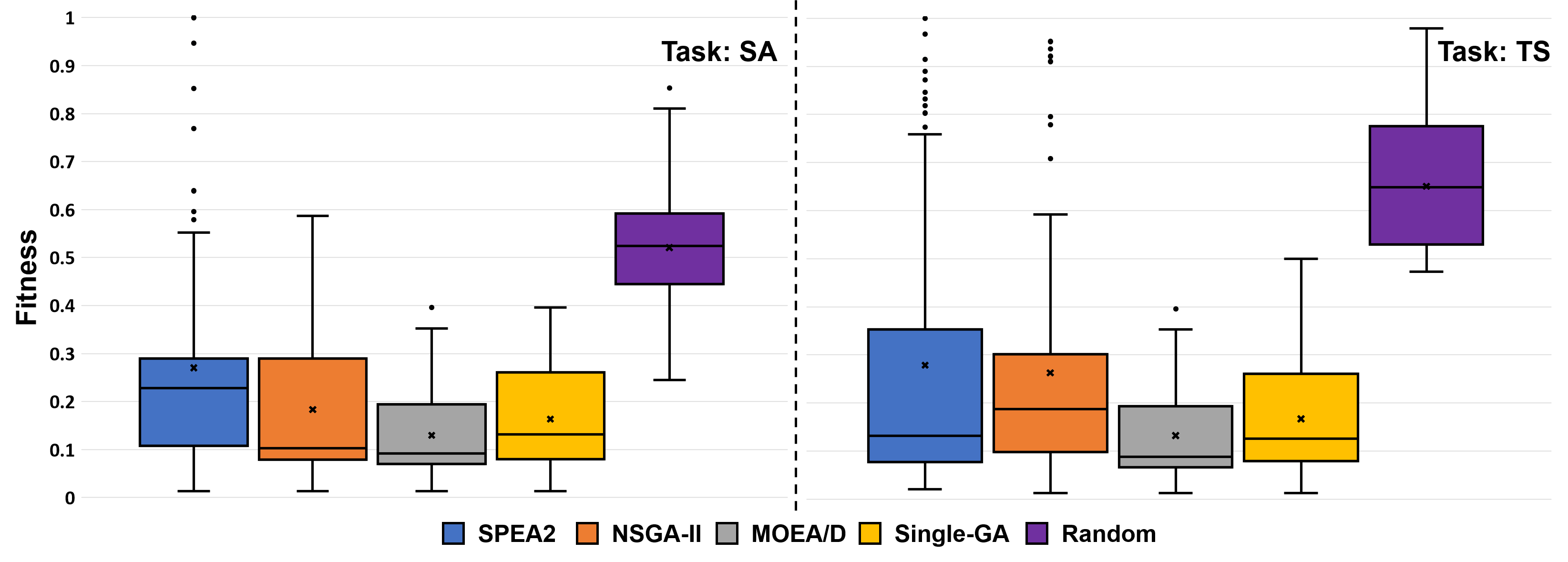}
    \caption{Fitness values from the experiment groups on evaluating the Robustness of SA and TS tasks for Llama-3.1 70B} \label{fig:RQ1-2}
    \vspace{5px}
\end{figure}

The \textbf{RQ1} aims to validate whether the suggested search algorithms significantly outperformed the baseline Random search method. Fig.~\ref{fig:RQ1-1} and \ref{fig:RQ1-2} illustrated the fitness values of the outcomes from the experiment group for targeting the SA and TS tasks for Gemini 1.5 Pro and Llama-3.1 70B models, respectively. In all cases of minimizing the fitness values for the two objectives, which are $(1 - Context\_ASR)$ and $C\_Token$, the Random search algorithm resulted in the highest fitness values. Notably, both the MOEA/D and Single-GA algorithms demonstrated consistent convergence of the fitness values compared to other search methods.

Furthermore, the varying levels of fitness value convergence among the search algorithms, as depicted in Fig.~\ref{fig:RQ1-1} and Fig.~\ref{fig:RQ1-2}, implemented in the LLMs suggest that Gemini 1.5 Pro is more resilient when dealing with robustness attacks on text data. Since the input text set remains consistent for both the Gemini and Llama model executions, meaning the $C\_token$ difference between the two LLMs is not significant, we found that the overall $Context\_ASR$ results in Gemini 1.5 are 8.93\% lower than those produced by the Llama 3.1 70B model.

First, we applied the Kruskal-Wallis (KW) test~\cite{kruskal1952use} to the fitness values of the experiment group. Table~\ref{tab:KW_F} presents the p-values obtained from the KW tests on corresponding tasks and LLMs. The KW results indicate that in all the cases of executing SA and TS tasks on the two LLMs, there exist statistically significant differences in distributions of fitness values across the experiment group.

Then, we conducted a pair-wise comparison of the proposed algorithms with the random search using the Mann-Whitney U (MWU) test~\cite{mcknight2010mann}. Due to the multiple comparisons, we adjusted $alpha$ to validate the p-value as 0.005 by using the Bonferroni adjustment~\cite{arcuri2011practical}. The details of the $\alpha$ adjustment are explained in Section~\ref{sec.expr.tv}. The test results of comparing the distribution of fitness values are explained in Table~\ref{tab:MWU}.

Each value in the Table~\ref{tab:MWU} content represents a p-value, with ``T" indicating that the p-value is less than 0.005. These p-values are derived from the MWU test, which compares the distribution of fitness values for outcome sets generated by different GA implementations and those produced by a random search algorithm. The results demonstrate statistically significant differences between the distributions of the search algorithms and the random methods in all cases.

\begin{table}[t!]
\centering
\caption{Kruskal-Wallis results of the experiment group on SA and TS tasks} \label{tab:KW_F}
\LARGE
\resizebox{0.7\textwidth}{!}{
\begin{tabular}{rcccc}
\hline \hline
\multicolumn{1}{r}{\textbf{Comparison Model}}         & \multicolumn{2}{c}{ \TBstrut \textbf{Gemini-1.5-Pro}} & \multicolumn{2}{c}{\textbf{Llama 3.1 70B}} \\ \hline
\multicolumn{1}{l}{}  \TBstrut        & \textbf{SA}          & \textbf{TS}          & \textbf{SA}          & \textbf{TS}         \\ \hline
\TBstrut \textbf{KW Test: P-value} & 3.84E-44             & 1.15E-75             & 2.16E-55             & 2.30E-64       \\ \hline \hline
\end{tabular}
}
\end{table}

\begin{table}[t!]
\centering
\caption{Mann-Whitney U test results of fitness values for implemented search algorithms compared with the Random search on SA and TS tasks in different LLMs} \label{tab:MWU}
\LARGE
\resizebox{0.87\textwidth}{!}{
\begin{tabular}{rcccc}
\hline \hline
\multicolumn{1}{r}{\textbf{Comparison Model}}         & \multicolumn{2}{c}{ \TBstrut \textbf{Gemini-1.5-Pro}} & \multicolumn{2}{c}{\textbf{Llama 3.1 70B}} \\ \hline
\multicolumn{1}{l}{}  \TBstrut        & \textbf{SA}          & \textbf{TS}          & \textbf{SA}          & \textbf{TS}         \\ \hline
\TBstrut \textbf{SPEA2 vs Random}     & 2.16E-05 (\textbf{T})             & 1.01E-08 (\textbf{T})             & 5.40E-27 (\textbf{T})             & 2.77E-24 (\textbf{T})            \\ \hline
\TBstrut \textbf{NSGA-II vs Random}   & 8.39E-19 (\textbf{T})             & 8.45E-32 (\textbf{T})             & 1.43E-32 (\textbf{T})             & 2.42E-37 (\textbf{T})            \\ \hline
\TBstrut \textbf{MOEA/D vs Random}    & 2.23E-23 (\textbf{T})             & 1.67E-47 (\textbf{T})             & 4.16E-44 (\textbf{T})             & 2.04E-48 (\textbf{T})            \\ \hline
\TBstrut \textbf{Single-GA vs Random} & 5.97E-31 (\textbf{T})             & 1.62E-48 (\textbf{T})             & 2.43E-42 (\textbf{T})             & 2.95E-48 (\textbf{T})           \\ \hline \hline
\end{tabular}
}
\end{table}

\begin{table}[]
\centering
\caption{Vargha-Delaney A effect size measure results of fitness values between the search algorithms with the Random search on SA and TS tasks in different LLMs}
\label{tab:VD-A}
\Large
\resizebox{0.8\textwidth}{!}{
\begin{tabular}{rcccc}
\hline \hline
\multicolumn{1}{r}{\TBstrut \textbf{Comparison Model}} & \multicolumn{2}{c}{\textbf{Gemini-1.5-Pro}} & \multicolumn{2}{c}{\textbf{Llama 3.1 70B}} \\ \hline
                               \TBstrut     & \textbf{SA}          & \textbf{TS}          & \textbf{SA}          & \textbf{TS}                 \\ \hline
\TBstrut \textbf{SPEA2 vs Random}                      & Medium (\textbf{\textless{}})       & Medium (\textbf{\textless{}})      & Large (\textbf{\textless{}})       & Large (\textbf{\textless{}})       \\ \hline
\TBstrut \textbf{NSGA-II vs Random}                 & Medium (\textbf{\textless{}})       & Large (\textbf{\textless{}})       & Large (\textbf{\textless{}})       & Large (\textbf{\textless{}})       \\ \hline
\TBstrut \textbf{MOEA/D vs Random}                & Large (\textbf{\textless{}})        & Large (\textbf{\textless{}})       & Large (\textbf{\textless{}})       & Large (\textbf{\textless{}})       \\ \hline
\TBstrut \textbf{Single-GA vs Random}           & Large (\textbf{\textless{}})        & Large (\textbf{\textless{}})       & Large (\textbf{\textless{}})       & Large (\textbf{\textless{}})       
\\ \hline \hline
\end{tabular}
}
\vspace{10px}
\end{table}

Finally, Table~\ref{tab:VD-A} presented the Vargha-Delaney A results for the same pairs in Table~\ref{tab:MWU}. The table provides the magnitude of $\hat{A}_{12}$ values for each analysis result, indicating the level of difference between the two data distributions~\cite{vargha2000critique}. The inequality signs in the table denote whether the $\hat{A}_{12}$ values are greater or less than 0.5. The $\hat{A}_{12}$ value over 0.5 means that the former algorithm achieved higher values in the comparison, while the $\hat{A}_{12}$ value less than 0.0 means that the latter algorithm returned higher values. 
The results revealed that the Random search demonstrated a significantly larger distribution of fitness values compared with the proposed search algorithms. Furthermore, the magnitude levels of the differences are mostly large except for the SPEA2 and NSGA-II results in Gemini-1.5 Pro.

\begin{tcolorbox}[colback=orange!5!white,colframe=orange!85!black]
The proposed search algorithms were significantly more effective for MR-based robustness testing optimization compared to the baseline Random search. Among the methods, the MOEA/D and Single-GA algorithms achieved the most converged fitness values.
\end{tcolorbox}

\textbf{RQ2.} We then conducted an in-detail investigation to specify the best search algorithm to facilitate the most optimized $MG$ scoping for the LLM robustness testing. 
We compared the HyperVolume indicator, fitness values, and execution time overhead of the proposed search algorithms.

\begin{figure}[t]
    \centering
    \includegraphics[width=0.98\textwidth, trim = 0cm 0cm 0cm 0cm,clip]{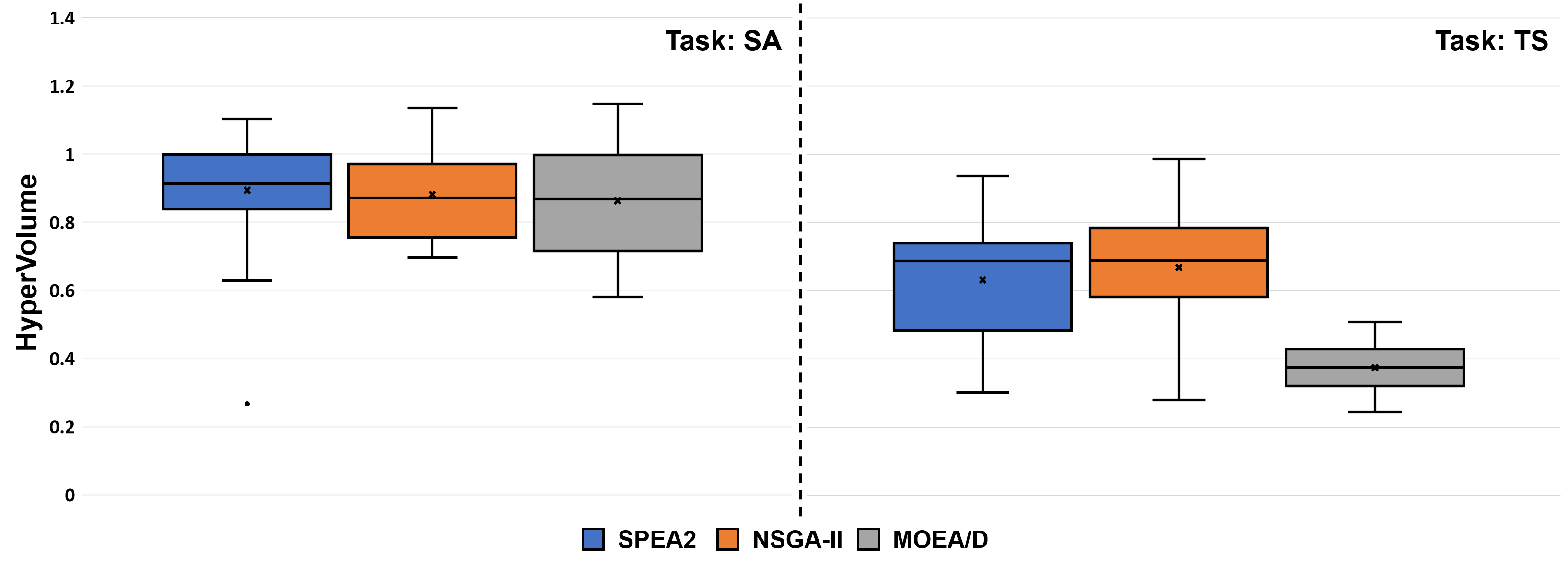}
    \caption{HyperVolume indicator evaluation results for multi-objective search algorithms of SA and TS tasks for Gemini 1.5 Pro} \label{fig:RQ2-1}
    \vspace{5px}
\end{figure}

\begin{figure}[t]
    \centering
    \includegraphics[width=0.98\textwidth, trim = 0cm 0cm 0cm 0cm,clip]{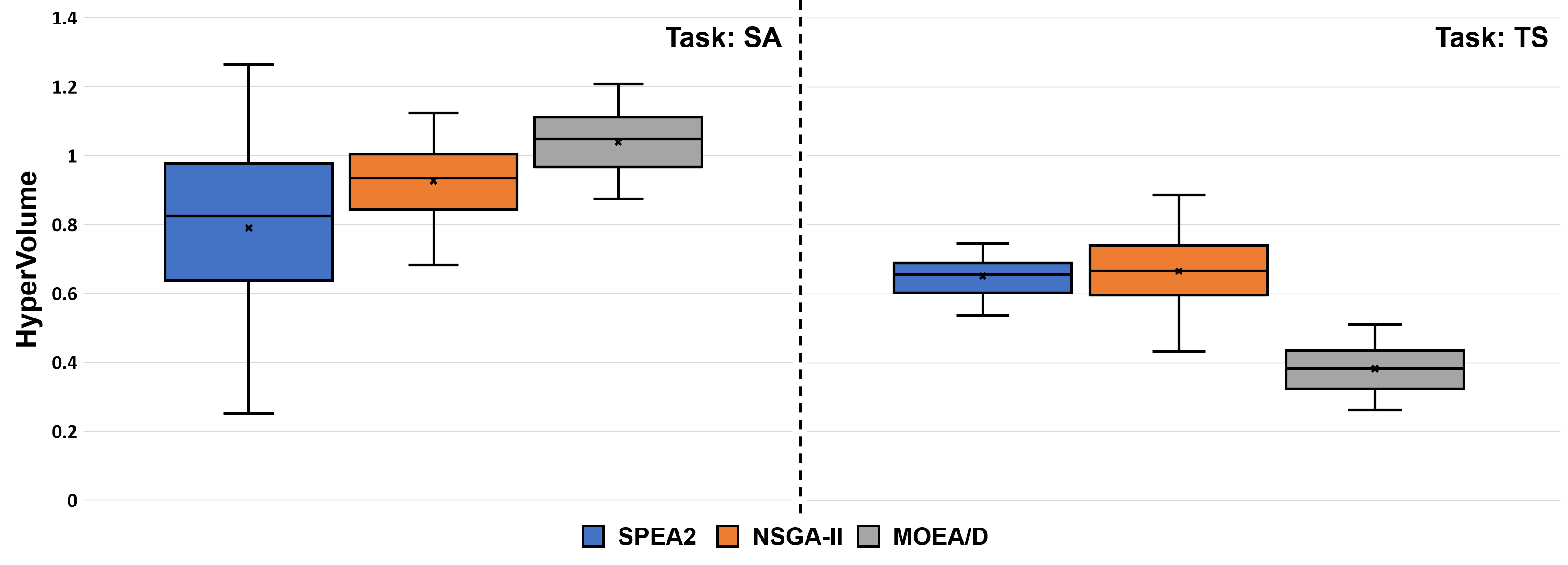}
    \caption{HyperVolume indicator evaluation results for multi-objective search algorithms of SA and TS tasks for Llama-3.1 70B} \label{fig:RQ2-2}
    \vspace{5px}
\end{figure}

\textbf{RQ2-1.} The \textbf{HyperVolume} (HV) indicator is one of the most widely used metrics in multi‐objective optimization, as it measures the volume in the objective space that is dominated by a set of non‐dominated solutions~\cite{zitzler2003performance}. In qualitative terms, a higher HV value means the solutions cover more of the Pareto front, indicating better trade‐off performance across the objectives. 
The HV analysis results are presented in Fig.~\ref{fig:RQ2-1} and Fig.~\ref{fig:RQ2-2}. In the SA task with smaller input token sets, the MOEA/D and NSGA-II achieved higher HVs when optimizing the MRs for robustness testing than those of SPEA2. However, in the TS task with larger input tokens, the MOEA/D exhibited the smallest HV values for both commercialized and local LLMs.

The results of the statistical investigation on the HV values for the multi-objective search algorithms are presented in Table~\ref{tab:KW_HV} and Table~\ref{tab:VD-A_HV}. The KW test results in Table~\ref{tab:KW_HV} indicate significant differences in the distributions of HV values for the SA tasks in the Gemini1.5 Pro and both tasks in the Llama 70B model. Furthermore, Dunn's test~\cite{dunn1964multiple} revealed that comparisons between the MOEA/D and the other two approaches consistently yielded significant p-values (e.g., 2.75E-07 with SPEA2 in SA - Gemini), all less than 0.005. In contrast, comparisons between NSGA-II and SPEA2 did not show any significant differences in HV distributions across all cases.

Table~\ref{tab:VD-A_HV} shows the in-detail analysis of the statistical effect sizes and scales. In SA tasks, the MOEA/D approach achieved similar or higher HV values than others while the NSGA-II and SPEA2 showed higher HVs than the MOEA/D.
Considering all the investigation results, we revealed that the NSGA-II algorithm consistently achieved high mean values HV scores across all tasks and target LLMs. Notably, in the TS task, the HV results for NSGA-II were nearly twice as high as those for the MOEA/D algorithm. 
This suggests that MOEA/D can specify the optimal Pareto Front space with smaller token inputs (e.g., SA task). Still, the approach can identify a few standout solutions (i.e., high fitness on the objectives) but cannot spread its solutions across the Pareto Front space, leading to the return of small and (local) optimal solution sets in assessing LLMs on large text tokens, including TS task. 

\begin{table}[t!]
\centering
\caption{Kruskal-Wallis test results of HV values for the multi-objective search algorithms on SA and TS tasks in different LLMs} \label{tab:KW_HV}
\LARGE
\resizebox{0.78\textwidth}{!}{
\begin{tabular}{rcccc}
\hline \hline
\textbf{Comparison Model}       & \multicolumn{2}{c}{ \TBstrut \textbf{Gemini-1.5-Pro}} & \multicolumn{2}{c}{\textbf{Llama 3.1 70B}} \\ \hline
\multicolumn{1}{l}{}  \TBstrut        & \textbf{SA}          & \textbf{TS}          & \textbf{SA}          & \textbf{TS}         \\ \hline
\TBstrut \textbf{KW Test: P-value}                                                                                 & 7.50E-10                                                                                                      & 0.428       & 2.23E-13                                                                                                      & 1.92E-06                                                                                                \\
\TBstrut \textbf{Dunn's Test: Significant Group} & MOEA/D   & -            & MOEA/D & MOEA/D
   \\ \hline \hline
\end{tabular}
}
\end{table}

\begin{table}[]
\centering
\caption{Vargha-Delaney A effect size measure results for HV values between the MOEA/D and other multi-objective algorithms on SA and TS tasks in different LLMs}
\label{tab:VD-A_HV}
\Large
\resizebox{0.75\textwidth}{!}{
\begin{tabular}{rcccc}
\hline \hline
\multicolumn{1}{r}{\TBstrut \textbf{Comparison Model}} & \multicolumn{2}{c}{\textbf{Gemini-1.5-Pro}} & \multicolumn{2}{c}{\textbf{Llama 3.1 70B}} \\ \hline
                               \TBstrut     & \textbf{SA}          & \textbf{TS}          & \textbf{SA}          & \textbf{TS}                 \\ \hline
\TBstrut \textbf{SPEA2 vs MOEA/D}                      & Small (\textbf{\textgreater{}})       & Large (\textbf{\textgreater{}})      & Large (\textbf{\textless{}})       & Large (\textbf{\textgreater{}})       \\ \hline
\TBstrut \textbf{NSGA-II vs MOEA/D}                 & Negligible (\textbf{\textgreater{}})       & Large (\textbf{\textgreater{}})       & Large (\textbf{\textless{}})       & Large (\textbf{\textgreater{}})       \\ \hline \hline
\end{tabular}
}
\vspace{10px}
\end{table}

In addition to the \textbf{Fitness} values in the outcomes described in Fig.~\ref{fig:RQ1-1} and Fig.~\ref{fig:RQ1-2}, we conducted further statistical analysis to identify the best search algorithm that can provide the most optimized $MG$ in the LLM robustness testing. Table~\ref{tab:KW_F2} describes the KW and Dunn's test results for the fitness values using the four proposed search algorithms. The results of the KW test indicated significant differences in the fitness distributions across all cases. Furthermore, we observed that the MOEA/D approach consistently exhibited significantly different distributions than the other methods in all task executions across different LLMs.

Based on this finding, we conducted the VD-A test to compare the MOEA/D with other search algorithms. Table~\ref{tab:VD-A_F} shows that the fitness values of the MOEA/D algorithm are lower than those of all other search algorithms in every comparison case. This indicates that the MOEA/D algorithm achieves the best convergence, meaning it effectively minimizes fitness values in the search for the optimal $MG$, outperforming other search-based approaches. The Single-GA exhibited the most similar performance in fitness minimization to the MOEA/D.


\begin{table}[t!]
\centering
\caption{Kruskal-Wallis test results of fitness values between the search algorithms on SA and TS tasks in different LLMs} \label{tab:KW_F2}
\LARGE
\resizebox{0.84\textwidth}{!}{
\begin{tabular}{rcccc}
\hline \hline
\textbf{Comparison Model}       & \multicolumn{2}{c}{ \TBstrut \textbf{Gemini-1.5-Pro}} & \multicolumn{2}{c}{\textbf{Llama 3.1 70B}} \\ \hline
\multicolumn{1}{l}{}  \TBstrut        & \textbf{SA}          & \textbf{TS}          & \textbf{SA}          & \textbf{TS}         \\ \hline
\TBstrut \textbf{KW Test: P-value}                                                                                 & 2.11E-20                                                                                                      & 2.06E-33       & 9.17E-09                                                                                                      & 4.63E-09                                                                                               \\
\TBstrut \textbf{Dunn's Test: Significant Group} & MOEA/D   & MOEA/D           & MOEA/D & MOEA/D
   \\ \hline \hline
\end{tabular}
}
\end{table}

\begin{table}[]
\centering
\caption{Vargha-Delaney A effect size measure results for fitness values between the MOEA/D and other search algorithms on SA and TS tasks in different LLMs}
\label{tab:VD-A_F}
\Large
\resizebox{0.82\textwidth}{!}{
\begin{tabular}{rcccc}
\hline \hline
\multicolumn{1}{r}{\TBstrut \textbf{Comparison Model}} & \multicolumn{2}{c}{\textbf{Gemini-1.5-Pro}} & \multicolumn{2}{c}{\textbf{Llama 3.1 70B}} \\ \hline
                               \TBstrut     & \textbf{SA}          & \textbf{TS}          & \textbf{SA}          & \textbf{TS}                 \\ \hline
\TBstrut \textbf{SPEA2 vs MOEA/D}                      & Large (\textbf{\textgreater{}})       & Large (\textbf{\textgreater{}})      & Medium (\textbf{\textgreater{}})       & Medium (\textbf{\textgreater{}})       \\ \hline
\TBstrut \textbf{NSGA-II vs MOEA/D}                 & Negligible (\textbf{\textgreater{}})       & Large (\textbf{\textgreater{}})       & Medium (\textbf{\textgreater{}})       & Small (\textbf{\textgreater{}})       \\ \hline
\TBstrut \textbf{Single-GA vs MOEA/D}                 & Negligible (\textbf{\textgreater{}})       & Large (\textbf{\textgreater{}})       & Small (\textbf{\textgreater{}})       & Small (\textbf{\textgreater{}})       \\ \hline \hline
\end{tabular}
}
\vspace{10px}
\end{table}

\begin{tcolorbox}[colback=orange!5!white,colframe=orange!85!black]
The MOEA/D algorithm generally performs well in optimizing the MRs for efficient robustness testing for several tasks with varying ranges of input tokens. Single-GA can be an alternative option due to the potential risks of local convergence of MOEA/D.
\end{tcolorbox}

\begin{figure}[t!]
\centering
    \includegraphics[width=0.85\textwidth, trim = 0cm 0cm 0cm 0cm,clip]{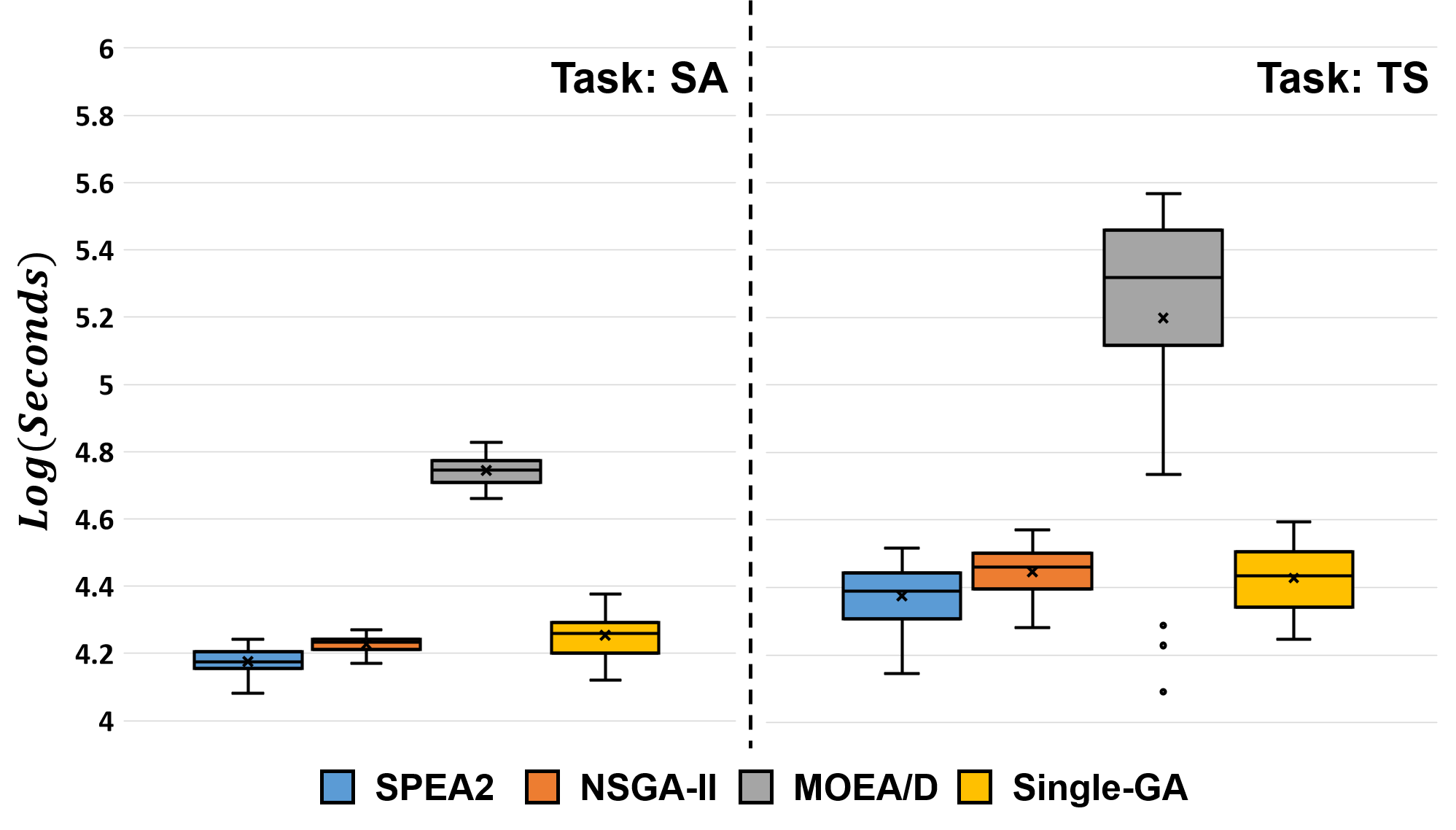}
    \caption{Execution overhead in log(seconds) for each search algorithm} \label{fig:RQ2-3}
    \vspace{10px}
\end{figure}

\textbf{RQ2-2.} Next, we examined the execution overhead of the proposed search algorithms to evaluate their practical feasibility. We recorded the execution times in seconds and visualized the logarithmic values of these times in Fig.~\ref{fig:RQ2-3}. While the NSGA-II, SPEA2, and Single-GA demonstrated similar execution times for the search algorithms, the MOEA/D exhibited significantly different execution overhead times. For instance, the MOEA/D algorithm for the SA tasks required between 46,125 and 67,102 seconds (approximately 12 to 18 hours) to complete 1,200 maximum iterations for one repetition. In contrast, other algorithms took between 13,214 and 23,194 seconds (around 3.5 to 6.4 hours) for the same configuration, indicating that the MOEA/D algorithm has about three times the execution overhead. This disparity becomes even more pronounced for the TS task, with the MOEA/D algorithm demanding nearly ten times more execution time than its counterparts.

\begin{tcolorbox}[colback=orange!5!white,colframe=orange!85!black]
The MOEA/D search algorithm achieves high capabilities of optimizing $MG$, but it may also incur significant execution time overhead. The other three search algorithms showed similar distributions in execution time.
\end{tcolorbox}

\begin{figure}[t]
\centering
    \includegraphics[width=0.78\textwidth, trim = 0cm 0cm 0cm 0cm,clip]{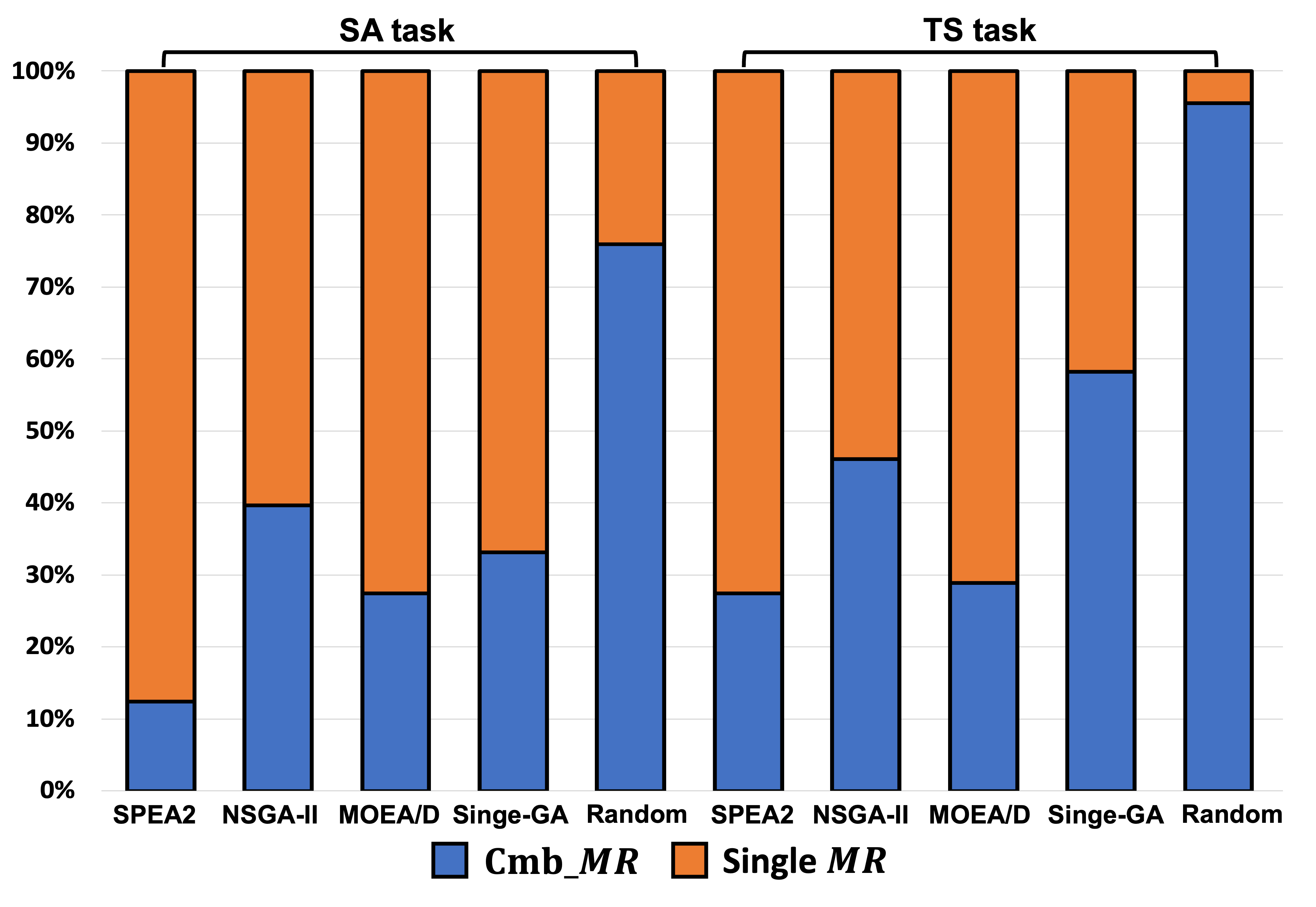}
    \caption{Distributions of single $MR$s and $Cmb\_MR$s in outcomes of search algorithms on Gemini 1.5 Pro} \label{fig:RQ3-1}
\end{figure}

\begin{figure}[t]
\centering
    \includegraphics[width=0.78\textwidth, trim = 0cm 0cm 0cm 0cm,clip]{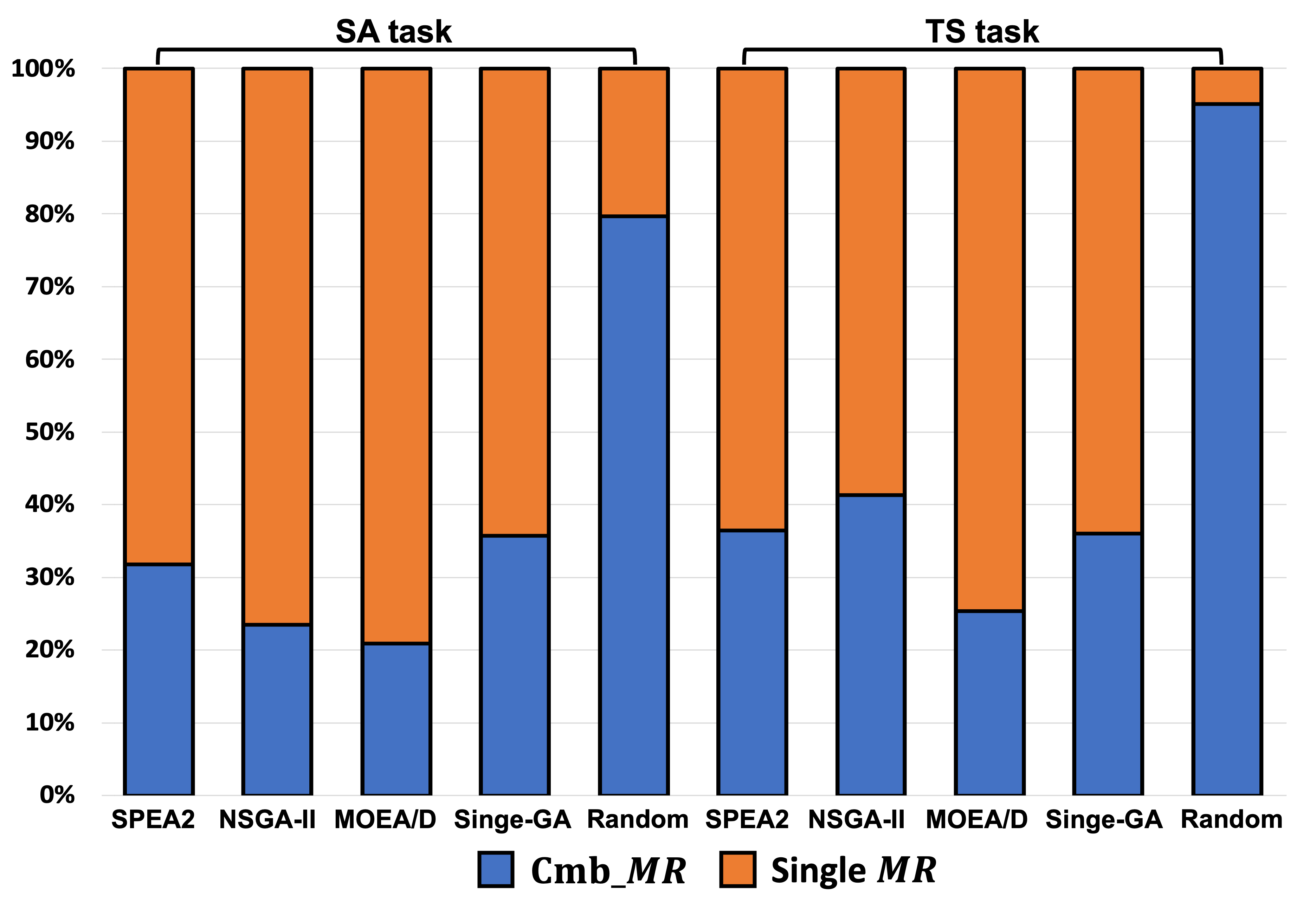}
    \caption{Distributions of single $MR$s and $Cmb\_MR$s in outcomes of search algorithms on Gemini 1.5 Pro} \label{fig:RQ3-2}
    \vspace{5px}
\end{figure}

\textbf{RQ3.} We have conducted an empirical analysis of the outcomes generated by all the search algorithms used in our experiment. First, we examined the distributions of single perturbation functions, referred to as single $MR$s, and those containing multiple perturbation functions, labeled as $Cmb\_MR$s, within the resulting $MG$s. Fig.~\ref{fig:RQ3-1} and Fig.~\ref{fig:RQ3-2} depict the distribution ratios for $Cmb\_MR$s and single $MR$s for each search approach while evaluating the SA and TS tasks on the Gemini 1.5 Pro and Llama 3.1 70B models, respectively.

For the evaluations conducted on the Gemini model, we found that approximately 44\% of the outcomes were $Cmb\_MR$s on average. Excluding the Random search, the proposed search algorithms generated $MG$s that contained an average of about 34\% $Cmb\_MR$s, with a minimum of 12.4\% and a maximum of 58.2\%. 
Similarly, in the robustness testing results for the Llama model, $ Cmb\_MR$s comprised 42.6\% of the overall outcomes. The outcomes produced by the proposed algorithms contained an average of 31.3\% $ Cmb\_MR$s, with a minimum of 20.9\% and a maximum of 41.3\%. 

Considering the exponentially large space of the $Cmb\_MR$s compared to that of the single $MR$s, we further examined the rationale to explain the distributions depicted in Fig.~\ref{fig:RQ3-1} and Fig.~\ref{fig:RQ3-2}. Our investigation revealed that specific single $MR$s exhibited the dominant effectiveness (i.e., failure revealing capabilities) in assessing the robustness of both commercialized and local LLMs with varying token sizes of input texts. Particularly, the character-level and graphical perturbation, \textit{L33TChanging()}, which changes ``meet" to ``m33t", and two word-level perturbations, \textit{AddRandomWord()} and \textit{SynonymReplacement()} are observed in the top-20 $MG$s from all search algorithms across all the execution cases. Due to the dominant effectiveness of the perturbation functions in confusing the LLMs' robustness, the search algorithms tended to converge on $MG$s composed of $Cmb\_MR$s containing these perturbation functions. Additionally, because the token cost of LLM executions, $C\_token$, is also an objective factor, we confirmed that the selection pressure favors a smaller number of independent $Cmb\_MR$s in a $MG$ with fewer executions of LLMs.

\begin{tcolorbox}[colback=orange!5!white,colframe=orange!85!black]
In our empirical analysis, we found specific dominant perturbation functions exhibiting outperforming effectiveness in testing the robustness of multiple LLMs on different tasks and input tokens. 
\end{tcolorbox}

\subsection{Threats to Validity} \label{sec.expr.tv}

\begin{figure}[t!]
    \centering
    \includegraphics[width=0.8\textwidth, trim = 0.3cm 0.3cm 0.3cm 0.3cm,clip]{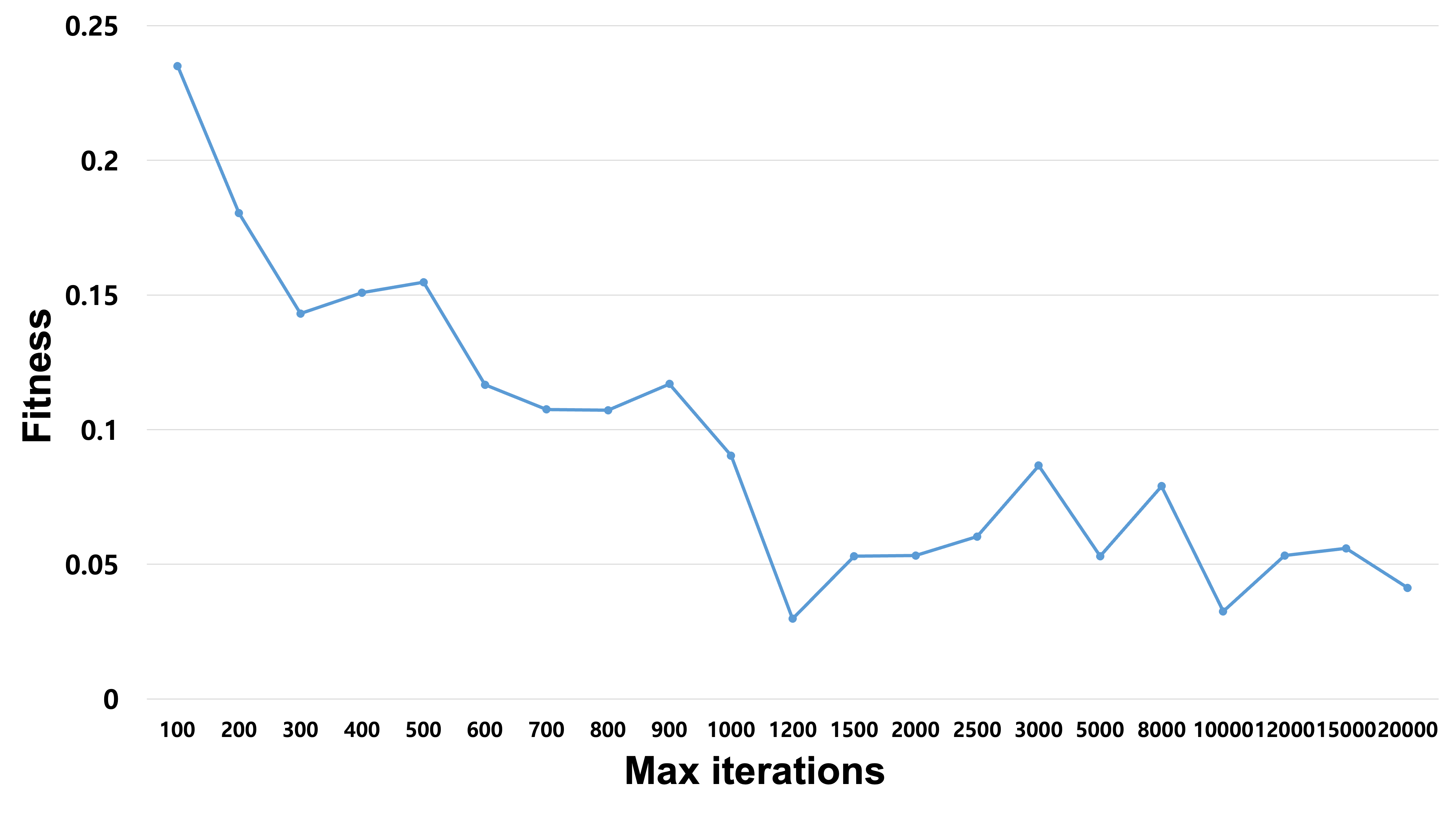}
    \caption{The pilot study results for Single-GA on the SA evaluation with different maximum iteration settings} 
    \label{fig:pilot}
    \vspace{5px}
\end{figure}

\textbf{Internal validity.} The hyperparameter settings described in Section~\ref{sec.expr.design} were configured based on the results of the pilot study using the Single-GA algorithm. Figure~\ref{fig:pilot} illustrates the average fitness values obtained from executing the Single-GA algorithm on SA tasks with the Gemini 1.5 pro model during this pilot study. The results indicate that using search algorithms to optimize the $MG$ with over 1,200 iterations produced similar average fitness values of approximately 0.05. Additionally, we found that the variance in fitness values from 1,500 to 20,000 iterations showed no significant differences compared to those at 1,200 iterations. Consequently, we set 1,200 iterations as the maximum iteration parameter for all search algorithms in our experiment. To determine the probabilities of search operations and the size of the $MG$, we referenced existing studies that employed GA and metamorphic testing techniques across various domains~\cite{cho2022automatic,ayerdi2021generating,applis2023searching} to establish conventionally acceptable probabilities and configurations.

\textbf{External validity.} We have executed the proposed search algorithms on Gemini 1.5 Pro, previously known as GooglePaLM~\cite{chowdhery2022palm}, and Llama 3.1 Pro. We have determined those models because they are the most suitable and free API-based and local LLMs to handle massive requests (approximately 3.5 million tokens for each repetition of every algorithm) during the optimization process. 
Additionally, we focused on finding solutions to the selection problem in MR-based LLM robustness evaluation. Unlike other studies that compare their adversarial attack methods with multiple target LLMs, we are concentrating on the feasibility validation of multiple search algorithms on target LLMs. We plan to extend the proposed approach to fine-tuned LLMs on specialized domains, such as Medical LLMs, and explore various quality issues.

\textbf{Construct validity.} As explained in Section~\ref{sec.appr.fitness}, there are two types of text perturbations: context-preserving and context-altering~\cite{hyun2024metal,liang2023holistic}. The perturbations shown in Fig.~\ref{fig:intro} are examples of context-preserving perturbations, while functions like \textit{ReplacingAntonymm()} or \textit{RemoveSentence()} are examples of context-altering perturbations. In our previous study, we applied different decision strategies for Pass/Fail checks (e.g., using``$=$" for context-preserving and ``$\neq$" for context-altering in LLM outcome comparison) and for context similarity (e.g., using $ContextSimilarity$ for context-preserving and $ContextDifference$ for context-altering in perturbation quality calculation) based on their perturbation types on single-level MRs~\cite{hyun2024metal}. However, in this study, we cannot determine the Pass/Fail and context comparison strategies in $Cmb\_MR$s that may contain both context-preserving and context-altering perturbations. Therefore, for the combined $Cmb\_MR$s, which include both types of perturbations, we initially decided the contextual similarity levels (e.g., strongly similar, weakly different, etc.) using thresholds from Google USE models~\cite{cer2018universal}. We then estimated which perturbation types the combined perturbations are more similar to. 

\textbf{Conclusion validity.} We conducted multiple comparisons of the three search-based algorithms in this experiment. To mitigate the increasing probability of Type I error in multiple statistical comparisons, we used Bonferroni adjustment to set the appropriate $\alpha$ values~\cite{arcuri2011practical}. Because we divided several experiment cases comparing outcomes from five search algorithms into two tasks, we calculated $\alpha$ for KW, Dunn's, and MWU tests as 0.005 by dividing 0.05 by 10 cases for each comparison case.

\section{Related Works}
\label{sec.related}

We investigated existing studies evaluating language models~\cite{hyun2024metal,polo2024tinybenchmarks,fan2023large, liu2020adversarial,jin2020bert,wang2023adversarial, zhu2023promptbench,zou2023universal} and studies applying search-based approaches to metamorphic testing for diverse target domains like cyber-physical systems~\cite{cho2022automatic, ayerdi2021generating} and AI-based software~\cite{applis2023searching}.

\textbf{Efficient Langauge Model Testing.} We examined recent studies that analyzed various types of language models to improve testing efficiency. Hyun et al.~\cite{hyun2024metal} proposed an automated testing framework, METAL, defining Metamorphic Relations (MRs) that can evaluate essential quality attributes on primary tasks in LLMs. Their study provides a set of MRs and automates the MR-based evaluation process. However, the study was limited to calculating the contribution of MRs in evaluating the effectiveness of specific tasks. 
Polo et al.~\cite{polo2024tinybenchmarks} recently suggested a sampling method called TinyBench to reduce LLM testing costs. They applied clustering and item response theory methods to curate small samples from the existing benchmark dataset. The main difference between our approach and TinyBench is the way we define individual items for sampling. Our study defined an MR-based $MG$ to optimize the subset of MRs, which can alleviate the oracle problem and can additionally reduce the true labeling cost. Because TinyBench defines a pair of text scenarios and correctness values extracted from the true labels and LLM results as a single item, we could not directly apply their approach to the MR set optimization problem context. Finally, we have investigated a recent survey paper exploring software engineering studies with LLMs by Fan et al.~\cite{fan2023large}. According to their survey, we have found that most software testing studies concentrated on applying LLMs to automate and minimize test suits for conventional software. 

There exist several studies exploring the automation of quality assessment for LLMs~\cite{liu2020adversarial,jin2020bert,wang2023adversarial} and revealing unreported quality issues in LLMs~\cite{zhu2023promptbench,zou2023universal} by generating adversarial attacks. 
Liu et al.~\cite{liu2020adversarial} trained an adversarial attack model that can randomly append spacing in the given text inputs, while Jin et al.~\cite{jin2020bert} developed an attack model facilitating the synonym substitutions in sentences. Wang et al.~\cite{wang2023adversarial} defined a demonstration in LLM execution and generated adversarial attacks focusing on security threats by the demonstration perturbations. Finally, Zhu et al.~\cite{zhu2023promptbench} and Zou et al. ~\cite{zou2023universal} formally defined the adversarial attack for input text and prompts and generated them by applying different levels of textual perturbations. However, the existing adversarial attack studies have focused on automating the specific attack generation for LLM inputs and only utilized the ASR method for analyzing the effectiveness of the results. While these studies have made significant progress in automating adversarial attacks for LLMs, their research scope did not cover the cost efficiency of generating perturbation attacks for LLM quality evaluation.

\textbf{Metamorphic Testing with Search-based Approaches.} Metamorphic Testing (MT) has been gaining significant attention for testing autonomous systems in various domains. Recent studies have proposed integrating MT with search-based heuristics, which can improve the cost efficiency and effectiveness of MRs in evaluating cyber-physical systems and AI-based software. Cho et al.~\cite{cho2022automatic} defined a set of GA operators for the continuous time-series data in autonomous driving LEGO vehicles. They applied the GA process to improve the effectiveness of pre-defined MRs in assessing the operation of autonomous vehicles. Similarly, Ayerdi et al.~\cite{ayerdi2021generating} proposed an automated MR generation approach based on the GA process. They generated several sets of MRs to evaluate the industrial elevator systems. The studies applied MT and GA for cyber-physical systems mainly addressed the oracle problem of continuous time-series data. Finally, Applis et al.~\cite{applis2023searching} provided a search-based MT approach to AI systems in a Java environment. They applied the transformation to the Java code and checked the robustness of AI codes by examining the differences in the results using Code2Vec. 

Our study proposed the search-based process to optimize the test space for evaluating LLM robustness. 
We further expanded the testing space by proposing $Cmb\_MR$s to improve effectiveness. However, existing studies applied MT and GA approaches to continuous time-series data, Java code, and benchmark data with true labels. 
In our study, we conducted extensive experiments on comparing the optimization performance of four well-known search algorithms and one baseline random search algorithm in the defined MR-based robustness test space for LLMs.

\section{Conclusion} \label{sec.conclusion}
In this study, we introduced a search-based testing process to select the optimal set of $Cmb\_MR$s for effective and efficient LLM robustness testing on Text-to-Text tasks. Our study formalized the efficiency and effectiveness measures of the LLM robustness evaluation. Also, we expanded the MR-based test space by covering the combinatorial application of different text perturbation functions.

We have developed four conventionally used search algorithms: Single-GA, NSGA-II, SPEA2, and MOEA/D as instances of the proposed search process. We executed the four search algorithms along with the baseline random search on two major text generation tasks on commercialized and local LLMs. Our investigation of the experiment results revealed that (1) MOEA/D achieved the best optimization fitness while Single-GA showed the most similar outcomes with MOEA/D and (2) the MOEA/D also requires the most significant amount of execution overhead compared with other algorithms. This indicates that the MOEA/D approach can evaluate the robustness of LLMs exhaustively for application scenarios that need a sophisticated level of fine-tuning and outcome accuracy, such as Medical LLMs in the digital healthcare domain. On the other hand, developers in small ventures can utilize the Single-GA approach for the cost-efficient assessment of personalized or fine-tuned LLMs. 

In our empirical analysis of the optimized $MG$, which is a subset of $ Set_Cmb_MR$s for assessing the robustness of LLMs, we identified three silver bullet perturbations: one graphical transformation and two word-level replacement and addition of semantically similar words. These perturbations demonstrate the dominant effectiveness in confusing different LLMs for all tasks and were consistently observed in all optimized outcomes generated by all the search algorithms in every iteration. 

To enhance the benefits of our approach, we plan to expand our approach to different fine-tuned LLMs from various domains. We will also investigate the extension of our search process to different quality factors, including explainability and fairness, and define the process to address the quality issues on LLMs by directly using the assessment outcomes.

\section*{Acknowledgements}
We thank Navaneeth Sivakumar for his assistance with developing the framework, conducting experiments, and improving the usability of the open repository.

\section*{Declarations}
\textbf{Funding:} The work has been supported by the Cyber Security Research Centre Limited, whose activities are partially funded by the Australian Government’s Cooperative Research Centres Program.
\newline
\\
\noindent
\textbf{Ethical Approval:} Not applicable.
\newline
\\
\noindent
\textbf{Informed Consent:} Not applicable.
\newline
\\
\noindent
\textbf{Author Contributions:} Sangwon Hyun - \textit{Conceptualization, Methodology, Software, Resources, Formal analysis and investigation, Writing - original draft preparation}, Shaukat Ali - \textit{Methodology, Validation, Writing - review and editing}, M. Ali Babar - \textit{Writing - review and editing, Supervision, Funding acquisition, Project administration}
\newline
\\
\noindent 
\textbf{Data Availability Statement:} The datasets and implemented codes for all the approaches are available at the following repository (https://zenodo.org/records/15205020).
\newline
\\
\noindent
\textbf{Conflict of Interest:} 
All authors certify that they have no affiliations with or involvement in any organization or entity with any financial or non-financial interest in the subject matter or materials discussed in this manuscript.
\newline
\\
\noindent
\textbf{Clinical Trial Number:} Not applicable.

\bibliographystyle{spbasic}
\bibliography{sample-base}

\end{document}